\documentclass[aps,prb,twocolumn,floatfix,showpacs,superscriptaddress]{revtex4-1}
\usepackage{graphicx}
\usepackage[colorlinks=true]{hyperref}
\usepackage{amssymb}
\usepackage{amsmath}
\newcommand{\diff}{{\rm{d}}}
\newcommand{\ket}[1]{\ensuremath{|#1\rangle}}
\newcommand{\bra}[1]{\ensuremath{\langle #1|}}

\begin{document}

\title{Electronic excitation spectra of the five-orbital Anderson impurity model: From the atomic limit to itinerant atomic magnetism}
\author{Li Huang}
\affiliation{Department of Physics, University of Fribourg, 1700 Fribourg, Switzerland}
\author{Tim O. Wehling}
\affiliation{Institute for Theoretical Physics, Bremen Center for Computational Materials Science, University of Bremen, 28359 Bremen, Germany}
\author{Philipp Werner}
\affiliation{Department of Physics, University of Fribourg, 1700 Fribourg, Switzerland}
\date{\today}

\begin{abstract}
We study the competition of Coulomb interaction and hybridization effects in the five-orbital Anderson impurity model by means of continuous time quantum Monte Carlo, exact diagonalization, and Hartree Fock calculations. The dependence of the electronic excitation spectra and thermodynamic ground state properties on hybridization strength and the form of the Coulomb interaction is systematically investigated for impurity occupation number $N \approx 6$. With increasing hybridization strength, a Kondo resonance emerges, broadens and merges with some of the upper and lower Hubbard peaks. Concomitantly, there is an increase of charge fluctuations at the impurity site. In contrast to the single orbital model, some atomic multiplet peaks and exchange split satellites persist despite strong charge fluctuations. We find that Hund's coupling leads to a state that may be characterized as an itinerant single atom magnet. As the filling is increased, the magnetic moment decreases, but the spin freezing phenomenon persists up to $N\approx 8$. When the hybridization is weak, the positions of atomic ionization peaks are rather sensitive to shifts of the impurity on-site energies. This allows to distinguish atomic ionization peaks from quasiparticle peaks or satellites in the electronic excitation spectra. On the methodological side we show that a comparison between the spectra obtained from Monte Carlo and exact diagonalization calculations is possible if the charge fluctuations are properly adjusted. 
\end{abstract}


\maketitle

\section{introduction\label{sec:intro}}
Strongly correlated materials with partially filled $d$ shells exhibit a range of interesting phenomena\cite{RevModPhys.70.1039,ADMA:ADMA201202018} which cannot be understood within the framework of static mean-field theories like density functional theory.\cite{PhysRev.136.B864,PhysRev.140.A1133} The photoemission spectra of these materials exhibit features which are reminiscent of atomic multiplets\cite{Groot200531} coexisting with quasiparticle bands. A formalism which captures the dual nature of itinerant and atomic like behavior is the dynamical mean-field theory\cite{RevModPhys.68.13,RevModPhys.78.865,GA:2004} which represents the solid by a self-consistently determined quantum impurity model. The impurity corresponds to a correlated multiorbital atom ($d$ shell) on a given lattice site, and fluctuates between different quantum states as electrons from neighboring sites hop in and out. Understanding how the spectral features of the isolated atom survive and change in this hybridized environment is important for the interpretation of calculated spectral functions and photoemission spectroscopy data. Similar questions arise in the study of adatom systems, where transition metal atoms are placed on different metallic surfaces and the spectral functions or transport properties are measured by photoemission spectroscopy or scanning tunneling microscopy.\cite{PhysRevLett.107.026801,PhysRevLett.110.136804,PhysRevLett.110.186404} In this context one also tries to understand the influence of the hybridization with the substrate on the local electronic properties.

Theoretically, such transition metal systems can be described in terms of the Anderson impurity model.\cite{PhysRev.124.41} The single orbital variant of this model is well understood and the dependence of excitation spectra on parameters like hybridization or temperature has been studied systematically.\cite{HewsonBook} However, much less is known about the multi-orbital case. Only multi-orbital impurities feature multiplet effects and it is still unclear how robust these effects are or how the multiplet features merge into broader Hubbard bands as the hybridization increases. 

Motivated by these considerations, we study the excitation spectra and valence fluctuations of a five-orbital Anderson impurity model over a range of hybridization strengths, from the atomic limit to the strongly hybridized case. The five-orbital Anderson impurity model describes transition metal adatom systems and has been considered in several recent publications. Correlation effects in this multiorbital system lead to nontrivial phenomena, such as orbitally controlled Kondo effects or Hund's exchange effects.\cite{PhysRevB.84.235110,PhysRevLett.110.186404,PhysRevB.85.085114,PhysRevB.80.155132,PhysRevB.81.115427,PhysRevB.85.161406,PhysRevLett.110.136804, Haule2009, Werner2012} In the present study, we focus on the occupation $N \approx 6$, which corresponds to an Fe$^{2+}$ impurity, and consider a flat density of states for the bath. The Fe$^{2+}$ configuration is very abundant and occurs in various systems including metalorganic molecules like Fe-porpherine,\cite{PhysRevLett.107.257202} Fe impurities embedded in topological insulators \cite{PhysRevLett.108.256811}, and Fe-pnictide and Fe-chalcogenide superconductors.\cite{ADMA:ADMA201202018} Fe in noble metal hosts is expected to have a $d$-electron occupation between $N=6$ and $N = 7$, which is likely closer to $N=6$ than to $N=7$ at least in the case of Fe on Ag surfaces.\cite{PhysRevLett.110.186404} Importantly, the $N=6$ configuration leads to the largest multiplet splittings in the atomic photoemission spectra in the late 3$d$ row. Thus, $N=6$ is particularly suitable for the study of the competition of atomic multiplet and charging features with hybridization effects, which is the major goal of this article. We will investigate the robustness of multiplet features against charge fluctuations and investigate if and how they evolve into Hubbard bands upon increasing hybridization.

We use three complementary techniques to solve the five-orbital Anderson impurity model: the continuous time hybridization expansion quantum Monte Carlo impurity solver (CT-HYB),\cite{PhysRevLett.97.076405,RevModPhys.83.349} exact diagonalization (ED)\cite{RevModPhys.68.13} as well as Hartree Fock (HF) approximation. Both ED and HF allow a direct calculation of the excitation spectra. However, ED can only handle few bath levels per orbital.\cite{ed:0953-8984-24-5-053201,PhysRevB.75.045125} HF can account for an arbitrary bath but neglects all dynamic correlation effects. In Monte Carlo simulations, on the other hand, the spectral function must be obtained using a numerically ill-conditioned analytical continuation.\cite{mem:1996} 

We address several points in the discussion of the five-orbital Anderson impurity model. First, it is not \emph{a priori} obvious how the bath parameters in ED should be chosen in order to enable a direct comparison with CT-HYB or HF. Here, we propose a strategy which is based on the measured charge fluctuations. Second, we show that the amount of charge fluctuations in the ground state as well as the excitation spectra are strongly affected by the type of Coulomb interaction matrix, even in the case of density-density interaction terms only. Using the fully rotationally invariant interaction (ED) and the density-density component of the rotationally invariant interaction (ED and CT-HYB), we then investigate the evolution of the impurity spectral function (including quasiparticle resonance peaks, satellite peaks, and multiplet features) as a function of the hybridization strength $V$. We demonstrate that the multiplet features can coexist with quasiparticles even in situations with strong charge fluctuations, where Hund's exchange $J$ can realize a regime of itinerant single atom magnets. We relate this observation to the spin-freezing phenomenon in multi-orbital lattice systems, and analyze the associated non-Fermi-liquid behavior as a function of filling. 

The paper is organized as follows: Section \ref{sec:model} defines the Anderson impurity model used in this study. Section \ref{sec:method} describes the computational details for the CT-HYB, ED, and HF methods. The results are presented in Section \ref{sec:results}, where we discuss the dependence of charge fluctuations and excitation spectra on the impurity hybridization and different forms of the Coulomb interaction matrix. Finally, concise summary and discussion are given in Section \ref{sec:conclusion}.

\section{model\label{sec:model}}
The five-orbital Anderson impurity model considered in this study is described by the Hamiltonian 
\begin{equation}
H_{\text{AIM}} = \sum_{k}\varepsilon_{k} c_{k}^{\dagger}c_{k} + \sum_{k,\alpha} (V_{k\alpha}c^{\dagger}_{k}d_{\alpha} + H.c) + H_{\text{loc}},
\label{eq:ham}
\end{equation}
and the local impurity term
\begin{equation}
\label{eq:hint}
H_{\text{loc}} = \sum_{\alpha}(\epsilon_{\alpha} - \mu) d^{\dagger}_{\alpha}d_{\alpha} 
+ \frac{1}{2} \sum_{\alpha\beta\gamma\delta} U_{\alpha\beta\gamma\delta}d^{\dagger}_{\alpha}d^{\dagger}_{\beta}d_{\gamma}d_{\delta}.
\end{equation}
Here, the $\alpha=(\sigma_\alpha, m_\alpha)$ denote combined spin and orbital indices, and the $d^{\dagger}_{\alpha}$ and $d_{\alpha}$ are the corresponding creation and annihilation operators for impurity electrons. The impurity level energy $\epsilon_{\alpha}$ is chosen to be zero (no crystal field splitting) and for the calculation of impurity spectral functions, the chemical potential $\mu$ is adjusted such that the total filling is $N = 6$. The local Coulomb interaction between the impurity electrons is parametrized by the average Coulomb interaction $U = 4.0$ eV and the Hund's exchange interaction $J = 1.0$ eV, which are reasonable values for typical transition metal atoms in a metallic environment.

There are two common prescriptions for deriving the Coulomb interaction matrix elements $U_{\alpha\beta\gamma\delta}$ from these parameters. A general and rotationally invariant form of this four-fermion Coulomb interaction can be obtained using the Slater parameters $F^{0} = U$, $F^{2} = 14/(1+0.625)J$, and $F^{4} = 0.625F^{2}$:
\begin{align}
U_{\alpha\beta\gamma\delta}&=\delta_{\sigma_\alpha,\sigma_\delta}\delta_{\sigma_\beta,\sigma_\gamma}\delta_{m_\alpha+m_\beta,m_\gamma+m_\delta}\nonumber\\
&\times\sum_{k=0}^4 c^k(m_\alpha;m_\delta)c^k(m_\gamma;m_\beta)  F^{k},
\label{eq:U_Slater}
\end{align}
where $c^k(m_\alpha;m_\delta)$ are Gaunt coefficients for angular momentum $l=2$, which are tabulated and explained in detail in Refs.~\onlinecite{Slater_Book,Eder_JuelichSchool12}. For the purpose of comparison, we will also consider a ``Slater-Kanamori" (S-K) type interaction of the form
\begin{align}
\label{eq:hint_sk}
H^\text{S-K}_\text{loc}&=\sum_{a,\sigma} (\epsilon_a-\mu)n_{a,\sigma} + \sum_{a} U n_{a,\uparrow} n_{a,\downarrow}\nonumber\\
&+\sum_{a>b,\sigma} \Big[U' n_{a,\sigma} n_{b,-\sigma} +  (U'-J) n_{a,\sigma}n_{b,\sigma}\Big]
\end{align}
with $U'=U-2J$. Since we neglect the spin-flip and pair-hopping terms in $H_{\text{loc}}^\text{S-K}$, this Hamiltonian is not rotationally invariant. It is worth mentioning that although Eq.~(\ref{eq:hint_sk}) may be used to approximate the Coulomb interaction in the $t_{2g}$ or $e_{g}$ manifold, it is not appropriate for the full $d$ shell. 

To mimic the effect of the lattice environment or substrate, the impurity is embedded in a sea of conduction electrons, which we label by a quantum number $k$. The creation and annihilation operators for the conduction electrons are $c^{\dagger}_{k}$ and $c_{k}$, and the energy level is $\varepsilon_k$. Finally, the coupling between the impurity electrons and the conduction electrons is parametrized by the hybridization strength $V_{k\alpha}$. The parameters $\varepsilon_k$ and $V_{k\alpha}$ define the hybridization function of the impurity model:
\begin{equation}
\label{eq:hyb}
\Delta_{\alpha}(i\omega_n) = \sum_{k} \frac{V^{*}_{k\alpha} V_{k\alpha}}{i\omega_n - \varepsilon_k} = V^{2}\int \diff\varepsilon \frac{\rho(\varepsilon)}{i\omega_n - \varepsilon}.
\end{equation}
In this paper, for the sake of simplicity, we assume that the hybridization function is diagonal and independent of the orbital. The hybridization strength $V$ is treated as an adjustable parameter and the bath density of states (DOS) $\rho(\varepsilon)$ is assumed to be normalized. We choose a flat DOS with full bandwidth $W = 20$ eV, namely, $\rho(\varepsilon) = 1/W$. We consider hybridization strengths between $V=0.0$, which represents the atomic limit, and $V=2.0$, which roughly corresponds to $3d$-transition metal impurities in noble metals.\cite{PhysRevB.85.085114}

\section{computational details\label{sec:method}}
To solve the model [see Eq.~(\ref{eq:ham})], we use three methods with complementary strengths and limitations. One approach is the exact diagonalization.\cite{PhysRevB.75.045125,ed:0953-8984-24-5-053201,RevModPhys.68.13} Because of the Hilbert space constraints, we can only treat a limited number of bath sites. We restrict the study of the five orbital impurity model to one bath level per orbital, which is a very rough representation of the bath with flat DOS. Nevertheless, this approach can give us valuable insights into the hybridization induced changes in the impurity spectra. The real frequency Green's functions can be directly obtained from the eigenstates $|\psi_n\rangle$ and eigenenergies $E_n$ using the Lehmann representation
\begin{align}
G(\omega+i0_+) &= \frac{1}{Z}\sum_{n,m}\frac{|\langle \psi_n|d^\dagger| \psi_m\rangle|^2 }{\omega+i0_+-E_n+E_m}\nonumber\\
&\times (e^{-E_m\beta}+e^{-E_n\beta}), 
\end{align}
and should be meaningful at least in the limit of weak hybridization. With a properly adjusted hybridization strength, the ED spectral function $A(\omega) = -\frac{1}{\pi} \text{Im} G(\omega+i0_+)$ also allows us to check and interpret the result obtained from the Monte Carlo calculation. 

The CT-HYB method is based on a diagrammatic expansion of the impurity partition function in powers of the hybridization function, and a stochastic sampling of collections of these strong-coupling diagrams.\cite{RevModPhys.83.349,PhysRevLett.97.076405} While this approach can handle the full Coulomb matrix, and has for example been used to study the multiorbital Kondo physics of Co adatoms,\cite{PhysRevB.85.085114} the numerical effort in the five orbital case is substantial.\cite{PhysRevB.80.235117} (For the Slater-Kanamori interaction, a significant speed-up is possible using the conserved quantities introduced in Ref.~\onlinecite{Parragh2012}.) On the other hand, a meaningful analysis of spectral features based on the maximum entropy (MaxEnt) analytical continuation\cite{mem:1996} of imaginary time Green's functions requires highly accurate Green's function data. Faced with this dilemma we chose for the present study to restrict the CT-HYB calculations to the density-density components of the interaction terms in Eqs.~(\ref{eq:hint}) and (\ref{eq:hint_sk}). With this approximation, the very efficient segment implementation of the CT-HYB algorithm becomes applicable.\cite{PhysRevLett.97.076405} We used about $10^9$ Monte Carlo samples per simulation to obtain accurate results, and employed the Legendre polynomial representation of $G(\tau)$ to filter the noise and suppress the numerical fluctuations.\cite{PhysRevB.84.075145} The CT-HYB spectral functions are then computed via the MaxEnt procedure, using a Gaussian default model with a smearing parameter $\sigma = 1.6$. The inverse temperature used in the CT-HYB calculations is set to $\beta = 40$, which corresponds to $T = 290$\ K, while all the ED calculations are performed at zero temperature.

Finally, we perform HF calculations, which help to distinguish static mean field from dynamic correlation effects and which can give valuable insights into ground state properties in terms of orbital and spin polarizations. The HF calculations are carried out in the wide band limit, where the self-consistency condition and the calculation of the impurity DOS can be implemented semi-analytically.\cite{PhysRev.124.41}

\section{results\label{sec:results}}
\subsection{Valence histograms and charge fluctuations}
\begin{figure}[hb!]
\centering
\includegraphics[width=\linewidth]{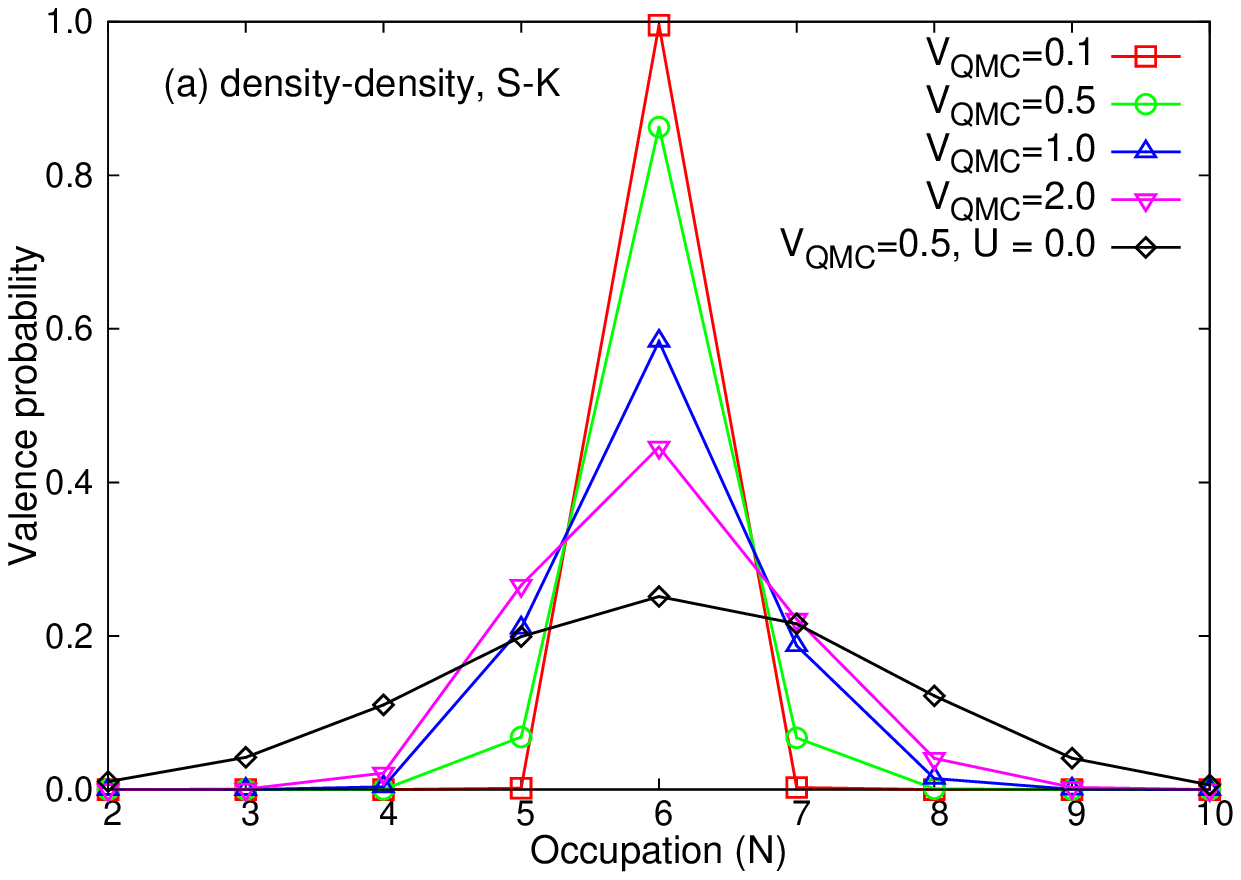}
\includegraphics[width=\linewidth]{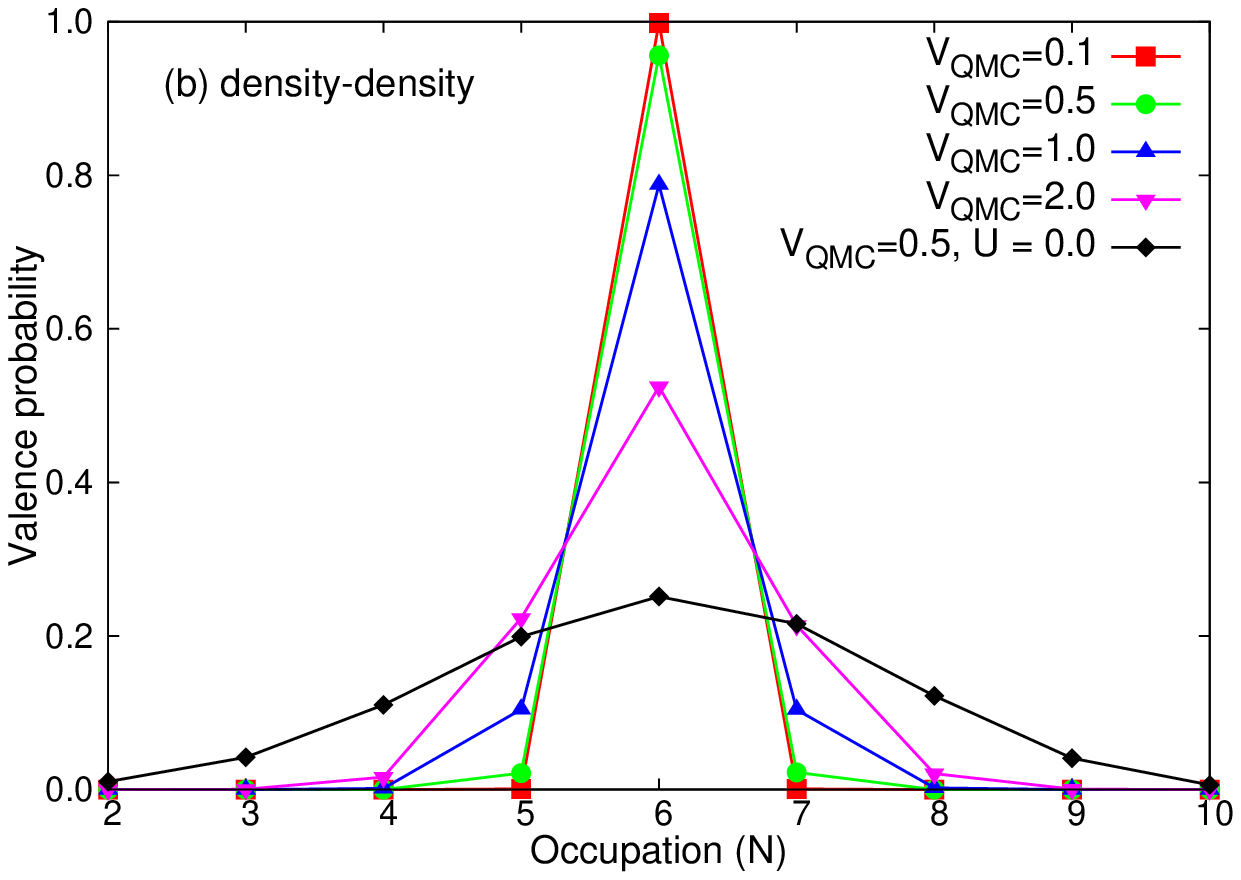}
\caption{(Color online) Valence probabilities for selected hybridization strengths $V_{\text{QMC}}$. The data are obtained by CT-HYB calculations. (a) Results for the Slater-Kanamori type density-density interaction. (b) Results for the general density-density interaction. For comparison, we also plot the valence histogram for the non-interaction case ($U = J = 0$) with $V_{\text{QMC}} = 0.5$ in panels (a) and (b). \label{fig:prob}}
\end{figure}

Since the bath representation in the ED and CT-HYB/HF calculations is very different, a comparison between ED and CT-HYB/HF spectra is only possible if the bath parameters $V_\text{ED}$ and $V_\text{QMC}$ are correctly matched. For this purpose, we propose to consider the average charge fluctuation $\sqrt{\langle N^2 \rangle - \langle N \rangle^2}$. In the atomic limit ($V=0$), the impurity is in the $N=6$ charge state and valence fluctuations are absent. As the coupling to the bath is turned on ($V > 0$), valence fluctuations occur and we obtain a probability distribution over different charge states, which is peaked at $N=6$. In Fig.~\ref{fig:prob} we show the CT-HYB results for the Slater-Kanamori and rotationally invariant interaction (both restricted to density-density components), for a range of hybridization strengths. For weak hybridization ($V_\text{QMC}<1.0$) one mainly observes fluctuations to the $N=5$ and $N=7$ states, while for $V_\text{QMC}=2.0$, also the $N=4$ and $N=8$ charge states gain significant weight. The valence histograms for the Slater-Kanamori and rotationally invariant cases are similar, with the latter exhibiting somewhat smaller charge fluctuations. The comparison to the histogram for the non-interacting impurity ($V_\text{QMC}=0.5$) shows that the interaction leads to a much more sharply peaked valence histogram, and thus (for the range of hybridization strengths considered in this study) to sizable electronic correlations.

Because the width of the valence histogram is an important number, which characterizes the state of the impurity, it is natural to choose the hybridization parameters $V_\text{ED}$ and $V_\text{QMC}$ such that the two calculations yield the same average charge fluctuation. For weak hybridization, the amount of charge fluctuations can be estimated perturbatively. To the lowest order in the hybridization, we find the following expression, which is analogous to Eq.~(8) of Ref.~\onlinecite{NozieresBlandin_1980}:
\begin{widetext}
\begin{align}
\label{eq:chgflct_pt}
\langle N^2 \rangle - \langle N \rangle^2=\sum_{k,\alpha} |V_{k\alpha}|^2\left((1-n_k)\left\langle d_\alpha^\dagger \left(\frac{1}{E_0-H_{\text{loc}}-\varepsilon_k}\right)^2 d_\alpha\right\rangle
 +n_k\left\langle d_\alpha \left(\frac{1}{E_0-H_{\text{loc}}+\varepsilon_k}\right)^2 d_\alpha^\dagger\right\rangle\right),
\end{align}
\end{widetext}
where $E_0$ is the ground state energy of the system without hybridization and $n_k$ is the occupation number of the bath state at $\varepsilon_k$. With the replacement $\sum_k \to \int\diff \varepsilon \rho(\varepsilon)$ and introducing the average coupling strength $V$ we have 
\begin{widetext}
\begin{align}
\label{eq:chgflct_pt2}
\langle N^2 \rangle - \langle N \rangle^2=|V|^2\sum_{\alpha}\int\diff\varepsilon\rho(\varepsilon) \left((1-n_\varepsilon)\left\langle d_\alpha^\dagger \left(\frac{1}{E_0-H_{\text{loc}}-\varepsilon}\right)^2 d_\alpha\right\rangle
+n_\varepsilon\left\langle d_\alpha \left(\frac{1}{E_0-H_{\text{loc}}+\varepsilon}\right)^2 d_\alpha^\dagger\right\rangle\right).
\end{align}
\end{widetext}
In the original model (which is solved by CT-HYB) we have $\rho(\varepsilon)=1/W$ for $-W/2<\varepsilon<W/2$ and thus
\begin{equation}
\langle N^2 \rangle - \langle N \rangle^2\approx \kappa\frac{2 |V|^2}{\Delta_{\text{Mott}}(\Delta_{\text{Mott}}+W)},
\label{eq:chgflct_pt3}
\end{equation}
where $\kappa$ is a numerical prefactor accounting for the orbital degeneracy and $\Delta_{\text{Mott}}/2$ is the energy required to excite the impurity from the $N=6$ ground state to an $N=5$ or $N=7$ state. Our model parameters lead to $W\gg \Delta_{\text{Mott}}$ and thus CT-HYB is expected to yield 
\begin{equation}
\langle N^2 \rangle - \langle N \rangle^2\approx \kappa\frac{2 |V|^2}{\Delta_{\text{Mott}}W}=\kappa\frac{2\text{Im}\Delta_\alpha(0)}{\pi \Delta_{\text{Mott}}}.
\label{eq:chgflct_pt3b}
\end{equation}
In the model studied by ED, the bath consists of one site per impurity orbital directly at the Fermi level and thus corresponds to the limit $W\to 0$:
\begin{equation}
\langle N^2 \rangle - \langle N \rangle^2\approx \kappa'\frac{2 |V|^2}{\Delta_{\text{Mott}}^2}.
\label{eq:chgflct_pt4}
\end{equation}

Eq.~(\ref{eq:chgflct_pt}) assumes a non-degenerate bath ground state, which is not the case in the limit $W\to 0$. Thus, the numerical prefactor $\kappa'$ in Eq.~(\ref{eq:chgflct_pt4}) will differ from the prefactor $\kappa$, which is found in the wide band case, Eq.~(\ref{eq:chgflct_pt3b}). Nevertheless, Eqs.~(\ref{eq:chgflct_pt3b}) and (\ref{eq:chgflct_pt4}) demonstrate that the amount of charge fluctuations $\sqrt{\langle N^2 \rangle - \langle N \rangle^2}$ should scale linearly with $V_{\text{QMC}}$ and $V_{\text{ED}}$. On the other hand, this perturbative treatment shows that it is natural to compare $|V_{\text{ED}}|^2/\Delta_{\text{Mott}}$ to the hybridization function $\text{Im}\Delta_\alpha(0)/\pi$ used in the CT-HYB calculations, or equivalently compare $|V_{\text{ED}}|/\Delta_{\text{Mott}}$ to $|V_{\text{QMC}}|/\sqrt{W\Delta_{\text{Mott}}}$. We expect that charge fluctuations in the discrete bath model solved by ED and the wide band model solved by CT-HYB will agree with each other up to a prefactor $\sqrt{\kappa'/\kappa}$ in this case.

\begin{figure}[t]
\centering
\includegraphics[width=\linewidth]{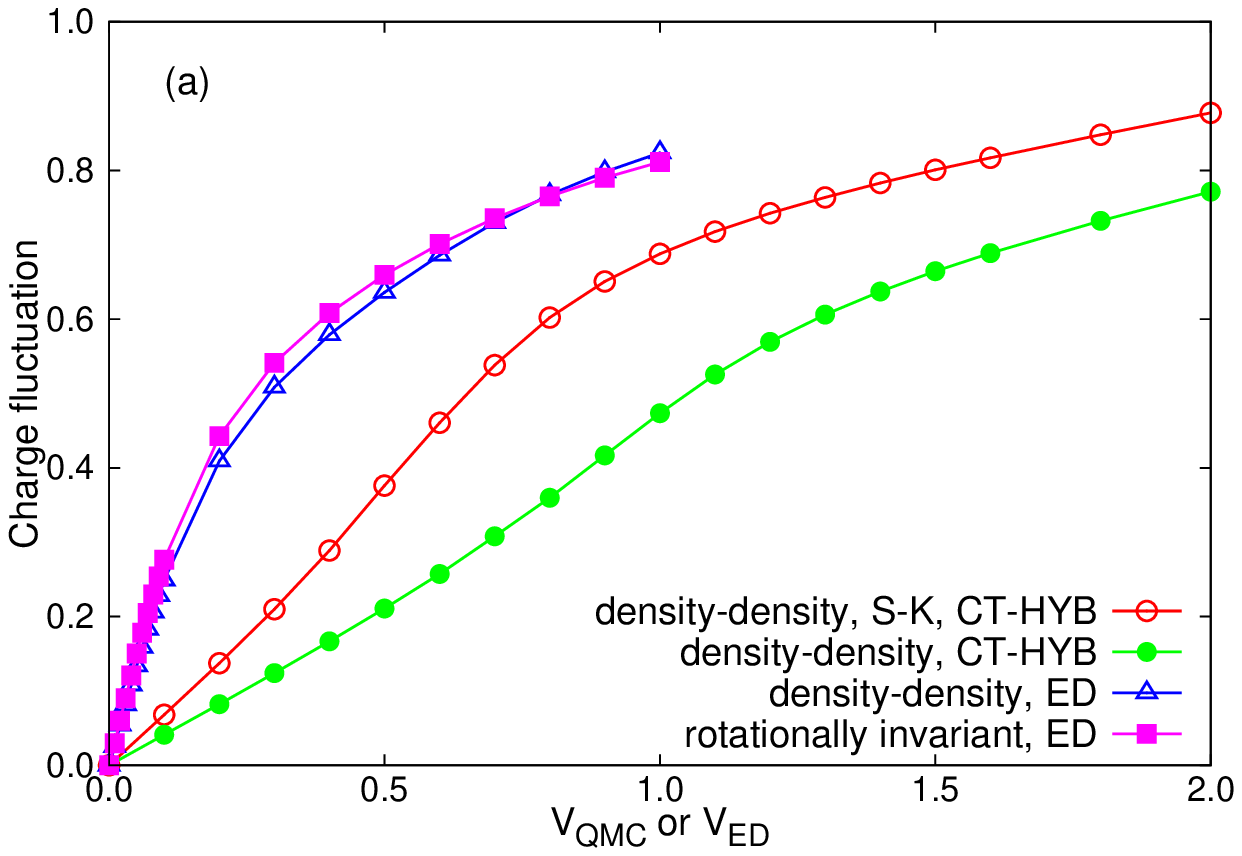}
\includegraphics[width=\linewidth]{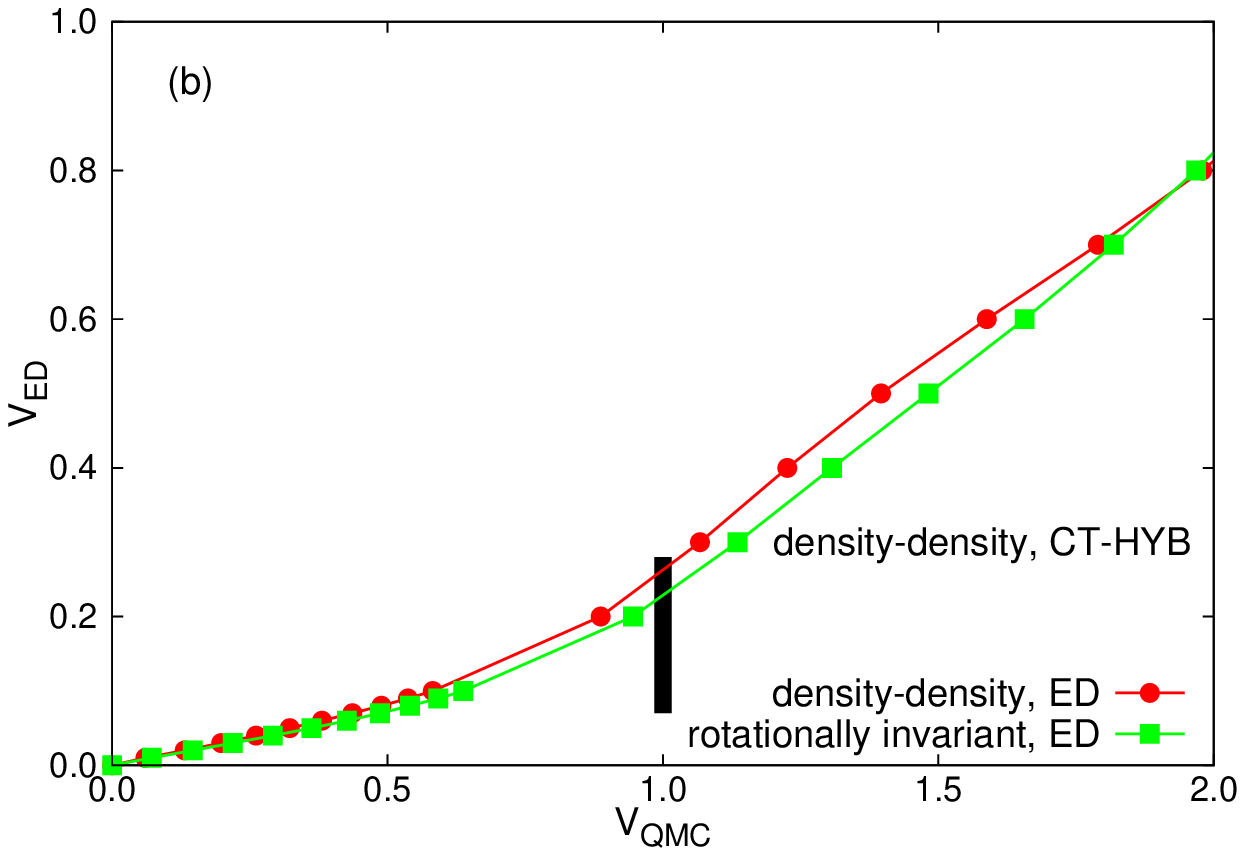}
\caption{(Color online) Upper panel: Charge fluctuations $\sqrt{\langle N^2 \rangle - \langle N \rangle^2}$ for different hybridization strengths $V_{\text{QMC}}$ and $V_{\text{ED}}$. Lower panel: Pairs of $V_{\text{ED}}$ and $V_{\text{QMC}}$ which give rise to the same charge fluctuations. The vertical black bar in $V_{\text{QMC}} = 1.0$ indicates the range of $V_{\text{ED}}$ obtained from the procedure suggested in Ref.~\onlinecite{PhysRevB.81.235125}. The fitting parameters are $20<\beta<100$ and $5 \leq n_c \leq 20$. \label{fig:flu}}
\end{figure}

In Fig.~\ref{fig:flu}(a) we plot the charge fluctuations as a function of $V_{\text{ED}}$ ($V_{\text{QMC}}$) for ED (CT-HYB), and different approximations of the Coulomb interaction matrix. This plot shows that for a given value of $V_{\text{ED}}$, the charge fluctuations obtained with the fully rotationally invariant Coulomb matrix, Eq.~(\ref{eq:U_Slater}), and only the density-density part of Eq.~(\ref{eq:U_Slater}) are comparable. Neglecting non-density-density terms in the Coulomb interaction matrix has obviously only a minor influence on the amount of charge fluctuations found at the impurity site. For a given value of $V_{\text{QMC}}$, in the CT-HYB case, the charge fluctuations are larger for the density-density part of the Slater-Kanamori Hamiltonian than for the density-density part of the rotationally invariant Coulomb matrix, Eq.~(\ref{eq:U_Slater}). The lower panel of Fig.~\ref{fig:flu} plots the pairs of hybridization strengths ($V_\text{QMC}$, $V_\text{ED}$) which correspond to the same value of the charge fluctuation. This graph can be used to find the suitable hybridization strength in an ED calculation, which allows a comparison to a given $V_{\text{QMC}}$ and vice versa.

It is worth noting that several schemes to obtain the ED hybridization parameters by fitting the full hybridization function [see Eq.~(\ref{eq:hyb})] of the Anderson impurity model on the imaginary or real frequency axis have been devised.\cite{PhysRevB.81.235125,PhysRevB.84.113112,PhysRevB.69.195105,liebsch:2012} These schemes typically rely on the definition of a weighting function which measures the distance between the full hybridization function and the ED one. For instance, a frequently employed approach [c.f. Ref.~\onlinecite{PhysRevB.81.235125}, Eqs.~(8) and (15a)] is to minimize the distance function
\begin{equation}
d=\sum_{n=0}^{n_c} |\Delta_{\text{full}}(i \omega_n)-\Delta_{\text{ED}}(i \omega_n)|,
\end{equation} 
for a given inverse temperature $\beta$ and cut-off Matsubara frequency $\omega_{n_c}$. For $V_{\text{QMC}}=1$, this procedure yields $0.07 < V_{\rm ED} < 0.27$ [visualized as a vertical bar in Fig.~\ref{fig:flu}(b)] for $20<\beta<100$ and $5 \leq n_c \leq 20$. The $V_{\text{ED}}$ obtained from our prescription of matching charge fluctuations in ED and QMC falls into this (rather wide) range, but does not involve any adjustable parameters like the cut-off $n_c$. In fact, as the results in the following subsections show, a factor 2 change in the hybridization parameter can lead to a qualitative change in the spectra.

\subsection{Influence of the Coulomb interaction matrix}
\begin{figure}[t]
\centering
\includegraphics[width=\linewidth]{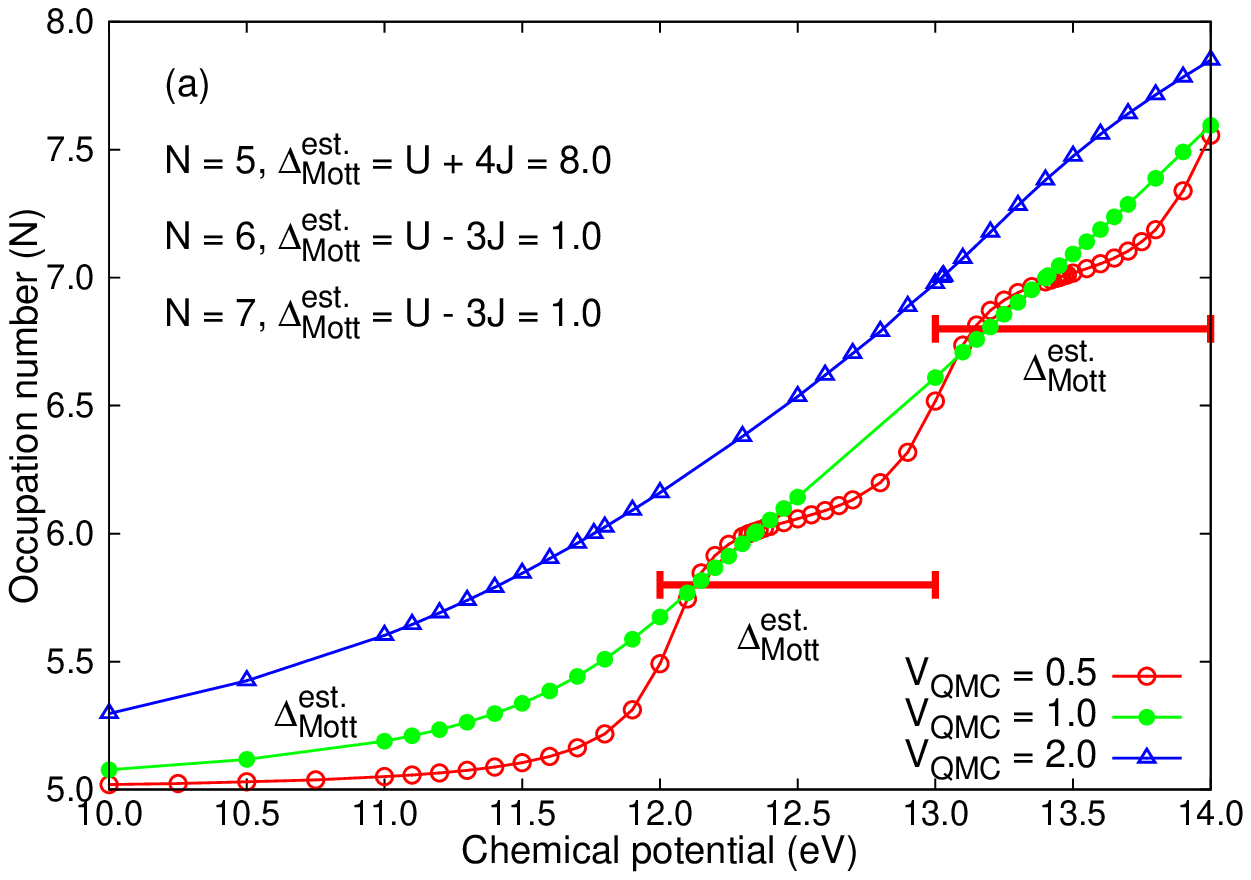}
\includegraphics[width=\linewidth]{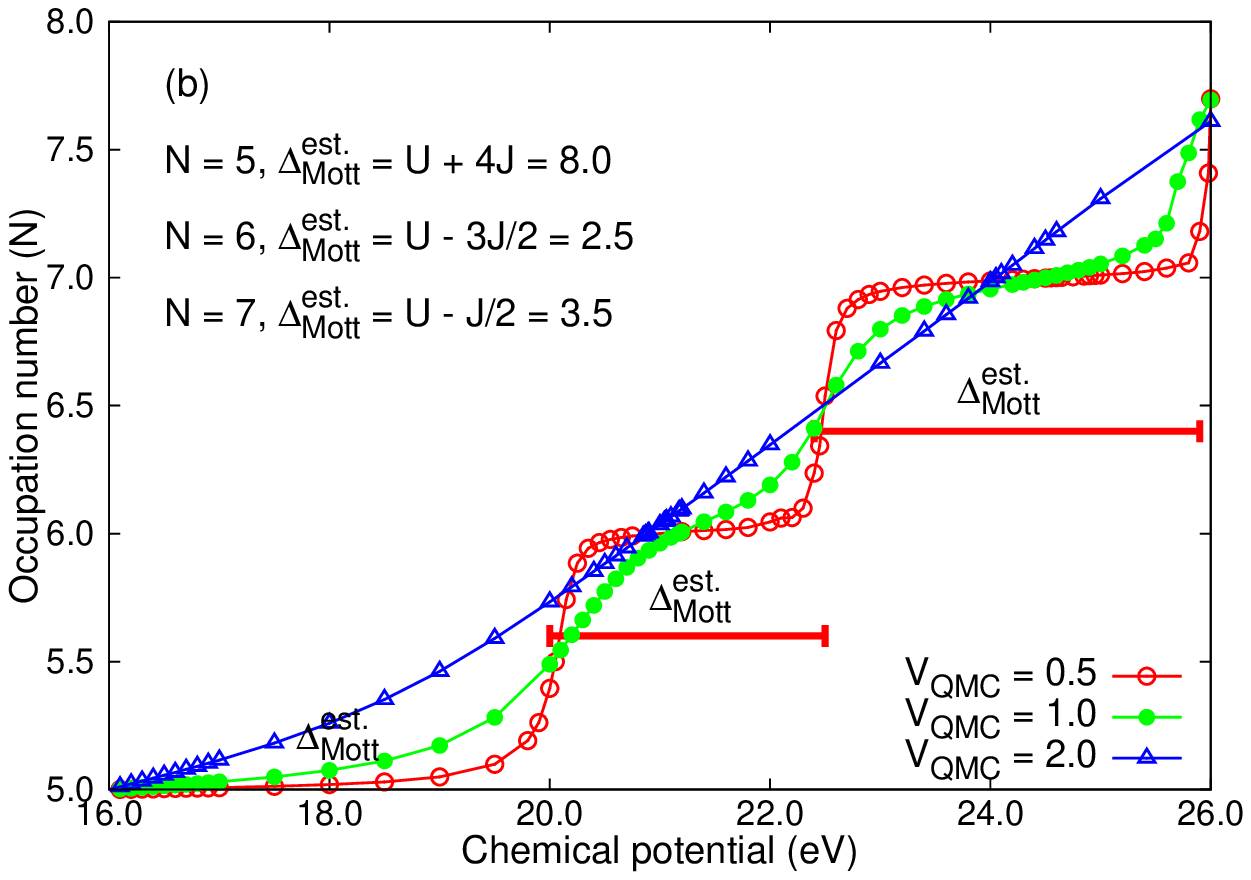}
\caption{(Color online) The total occupation number $N$ as a function of chemical potential $\mu$ and hybridization strength $V_{\text{QMC}}$. In the top panel, the Coulomb interaction matrix takes the Slater-Kanamori density-density form. In the bottom panel, it is extracted from a general Coulomb interaction matrix by keeping the density-density part. These data were obtained in CT-HYB simulations with $V_{\text{QMC}}$ = 0.5, 1.0, and 2.0, respectively. The red segments mark numerical estimates of the ``Mott plateau" $\Delta_{\text{Mott}}$. $\Delta^{\text{est.}}_{\text{Mott}}$ is the theoretically estimated value. \cite{PhysRevLett.110.186404} \label{fig:chem_occ}}
\end{figure}

\begin{figure}[t]
\centering
\includegraphics[width=\linewidth]{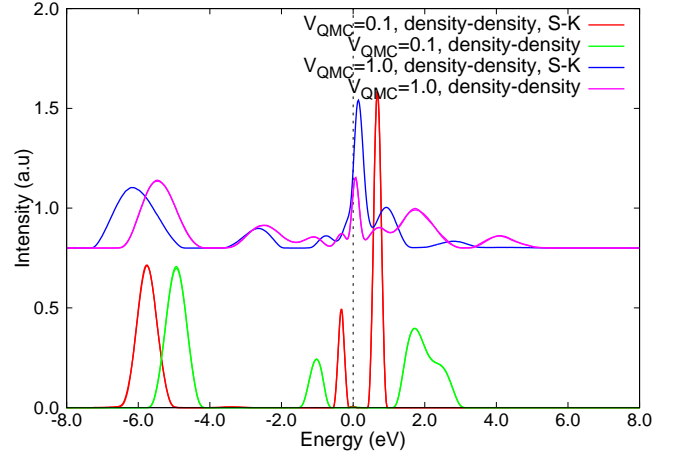}
\caption{(Color online) Spectral functions for $V_{\text{QMC}}$ = 0.1 and 1.0. The data are obtained by CT-HYB calculations and then post-processed using the MaxEnt method.\label{fig:ct_v01_v10}}
\end{figure}

\begin{figure}[t]
\centering
\includegraphics[width=0.97\linewidth]{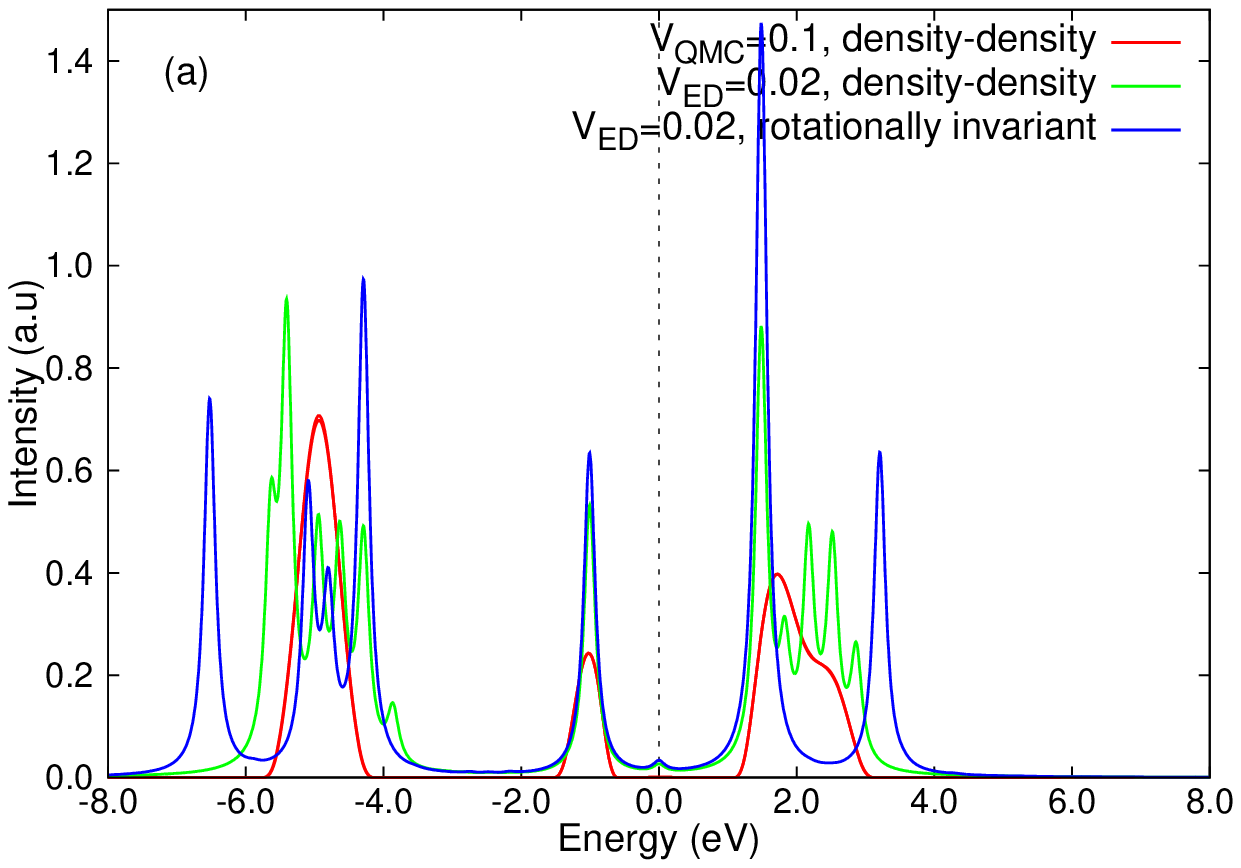}
\includegraphics[width=0.97\linewidth]{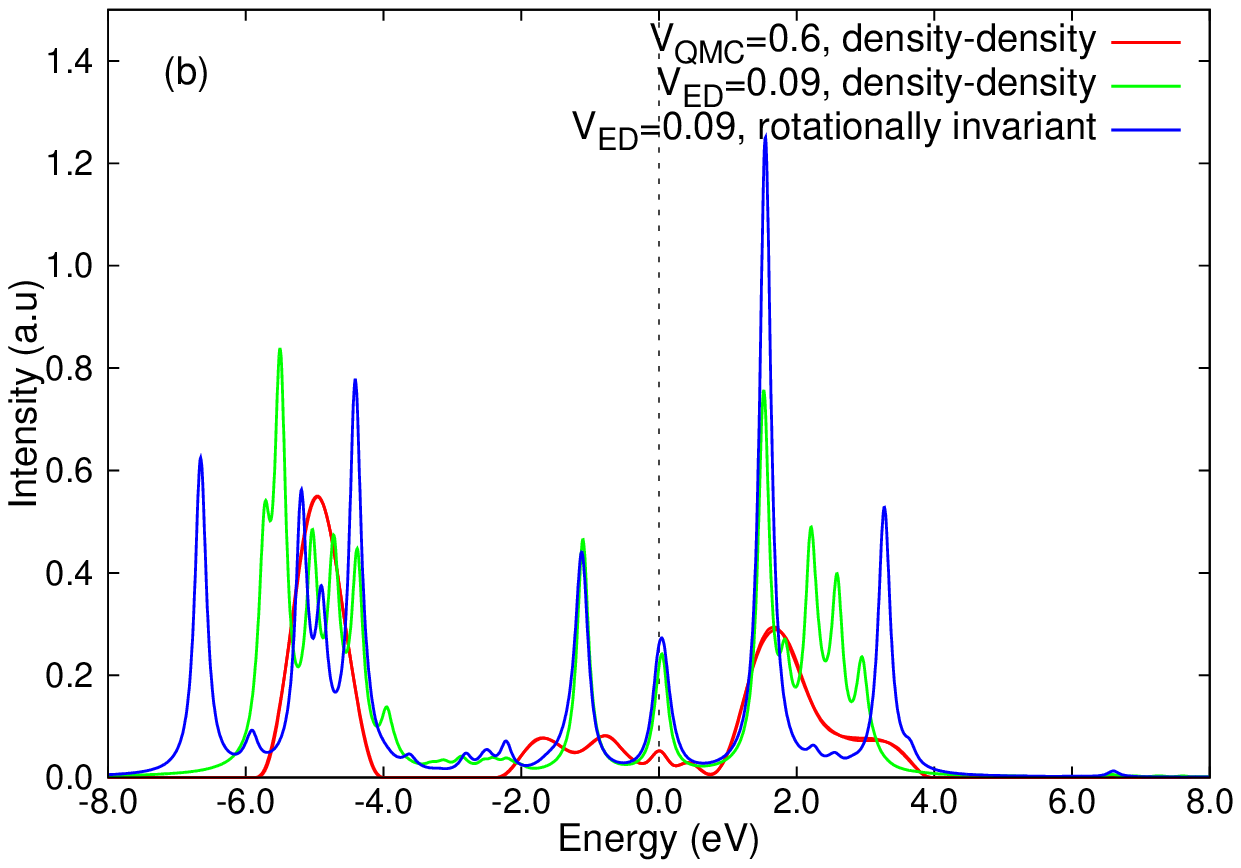}
\includegraphics[width=0.97\linewidth]{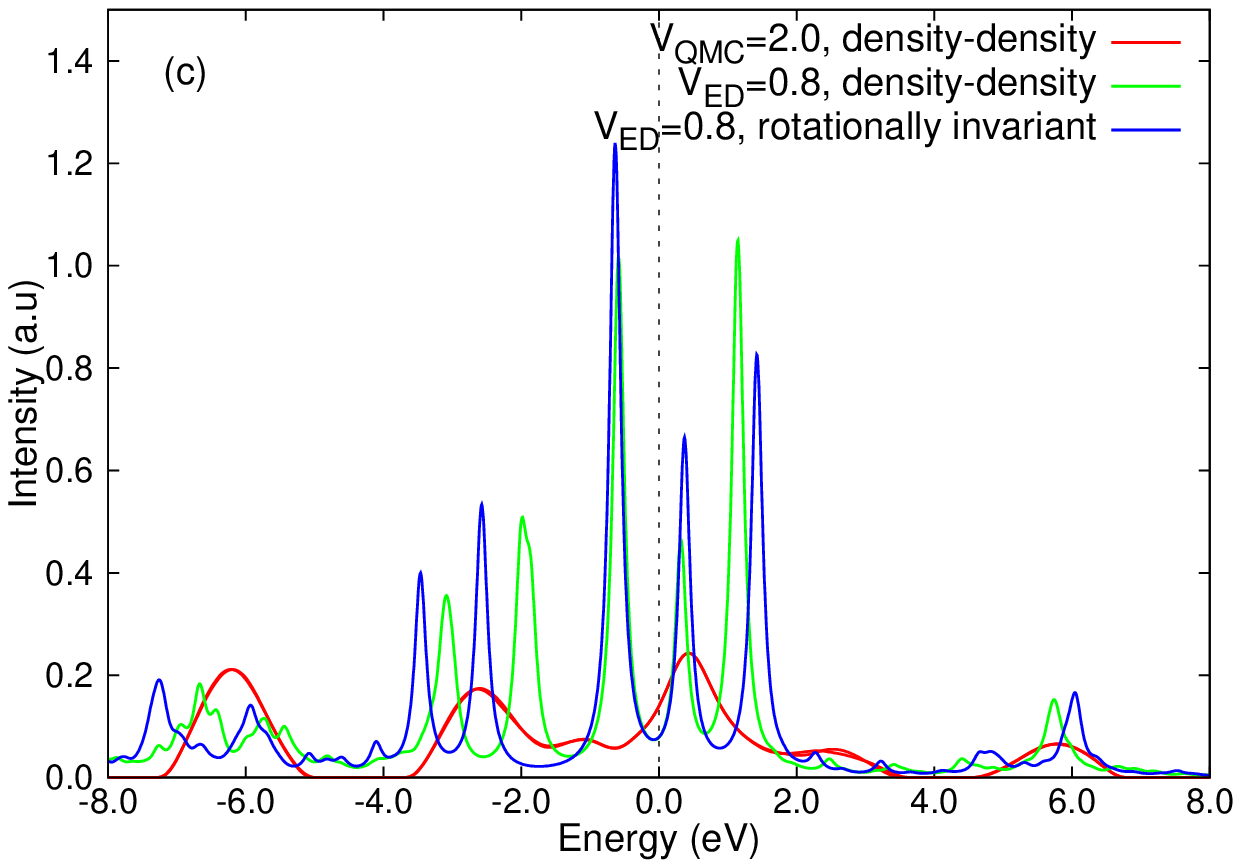}
\caption{(Color online) Comparison of spectral functions obtained by ED and CT-HYB calculations. (a) $V_{\text{ED}}$ = 0.02 and $V_{\text{QMC}}$ = 0.1. (b) $V_{\text{ED}}$ = 0.09 and $V_{\text{QMC}}$ = 0.6. (c) $V_{\text{ED}}$ = 0.8 and $V_{\text{QMC}}$ = 2.0. The charge fluctuations for the given ($V_{\text{ED}}$, $V_{\text{QMC}}$) pairs are comparable. \label{fig:ct_v01_ed_v00}}
\end{figure}

In a previous study of Co impurities in Cu hosts, it has been demonstrated that the approximations on the Coulomb interaction matrix may have a big influence on the calculated results.\cite{PhysRevB.80.155132} In the present study, while ED can treat the fully rotationally invariant Coulomb matrix, the QMC calculations are restricted to density-density interactions. Thus, we have to consider the effect of the interaction matrix used in the calculations. 

Irrespective of whether one treats all the interaction terms or just the density-density components, it is very important to start from the rotationally invariant form [see Eq.~(\ref{eq:U_Slater})] to construct the interaction matrix. We have already seen in Fig.~\ref{fig:flu} that the charge fluctuations in the Slater-Kanamori case are larger than in the rotationally invariant case. The reason for this difference is that the charge excitation gap in the weakly hybridized case is much smaller if the Slater-Kanamori interaction is used. To illustrate this, the CT-HYB results for the filling as a function of chemical potential $\mu$ are plotted in Fig.~\ref{fig:chem_occ}. The top panel shows CT-HYB results for the Slater-Kanamori interaction, and the bottom panel CT-HYB results for the rotationally invariant interaction (only the density-density components in both cases). If the hybridization strength is weak, plateaus appear around integer occupations. However, the $N=6$ plateau is seen to be significantly broader in the rotationally invariant calculations.

The width of the ``Mott plateau" can be roughly estimated by the formula $\Delta_\text{Mott}^{(N)}=E(N+1)+E(N-1)-2E(N)$, where $E(N)$ is the lowest atomic energy of the $N$ electron state.\cite{Werner2009} In the Slater-Kanamori case, this gives $\Delta_\text{Mott}^{(N=5)}=U+4J$ (half-filling), and $\Delta_\text{Mott}^{(N\ne 5)}=U-3J$. The estimate for the rotationally invariant interaction case is the same for $N=5$, but away from half-filling, the result differs: $\Delta_\text{Mott}^{(N=6,9)}\approx U-3J/2$ and $\Delta_\text{Mott}^{(N=7,8)}\approx U-J/2$.\cite{PhysRevLett.110.186404} Thus, for $N=6$ and the parameters $U=4.0$ eV and $J=1.0$ eV, one finds a gap of 1.0~eV in the Slater-Kanamori case and a gap of 2.5~eV in the rotationally invariant case, which is very close to the calculated widths of the Mott plateaus. Furthermore, as shown in Fig.~\ref{fig:ct_v01_v10} and Fig.~\ref{fig:ct_v01_ed_v00} the gaps in the electronic excitation spectra for small hybridization strengths are roughly consistent with this estimate. In the more strongly hybridized case ($V_\text{QMC}=1.0$) the quasiparticle peak in the Slater-Kanamori calculation becomes much more prominent than that in the rotationally invariant calculation (see Fig.~\ref{fig:ct_v01_v10}). These large qualitative and quantitative differences show that one cannot use the Slater-Kanamori form of the interaction in the five orbital case. For the rest of this paper we will therefore only discuss results for (the density-density component of) the rotationally invariant Coulomb interaction.  

As a test of the trustworthiness of the features in the MaxEnt spectral functions and to judge the effect of non-density-density Coulomb interaction terms, we compare in Fig.~\ref{fig:ct_v01_ed_v00} the CT-HYB and ED results for small, medium, and relatively large hybridizations. The hybridization strengths used in the CT-HYB and ED calculations have been adjusted according to the charge fluctuation prescription introduced in the previous subsection. Figure~\ref{fig:ct_v01_ed_v00}(a) resembles the atomic limit, Fig.~\ref{fig:ct_v01_ed_v00}(b) corresponds to a moderate hybridization strength, and Fig.~\ref{fig:ct_v01_ed_v00}(c) represents the strongly hybridized case.

In the ED case, we show spectral functions for both the fully rotationally invariant and the density-density interaction, obtained by truncating the rotationally invariant one. The former spectral function has less peaks because of the higher symmetry. The Mott gap is not affected by reducing the Coulomb interaction to density-density terms only. The degeneracy of the lowest $N=7$ multiplet is, however, reduced for the density-density interaction, which leads to a slight reduction of charge fluctuations in this case [see Fig.~\ref{fig:flu}(a)].

For the comparison of the ED and the CT-HYB MaxEnt spectrum, we should consider the density-density result. Overall, the agreement with the CT-HYB spectral function is satisfactory. In particular, for the small hybridization strength, neither of the spectral functions features a quasiparticle peak, while for the stronger hybridization, the weight of the quasiparticle peak is roughly consistent. Also, the energies of the main spectral features in the MaxEnt spectrum seem to coincide with multiplet peaks or satellites found in the ED calculations. For the fully rotationally invariant Coulomb interaction, the spectral peak at $\omega=-1$\,eV in Fig.~\ref{fig:ct_v01_ed_v00}(a) can be traced back to an excitation from the $N=6,L=2,S=2$ impurity ground state to an $N=5,L=0,S=5/2$ high-spin final state, while the excitations at $\omega<-4$\,eV belong to $N=5, S=3/2$ low-spin final states. The density-density type interactions break spin and orbital rotation symmetry. Thus, the final states are not necessarily eigenstates of orbital $L$ and spin angular momentum $S$ any more. While the spectral feature at $\omega=-1$\,eV can still be traced back to the $N=5,L=0,S=5/2$ final state, such an assignment is no more possible for the excitations at $\omega<-4$\,eV.

Judging from these results, it seems not unreasonable to systematically study the evolution of the MaxEnt spectral functions with increasing hybridization strength. 

\subsection{Evolution of the spectral functions}
\begin{figure}[t]
\centering
\includegraphics[width=\linewidth]{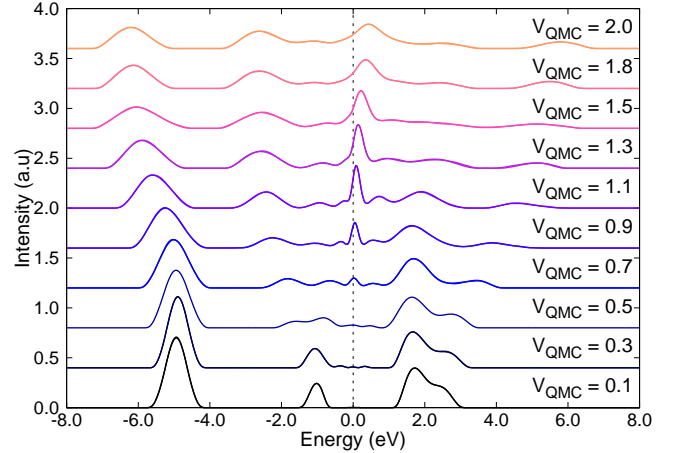}
\caption{(Color online) Calculated spectral functions at different hybridization strengths $V_{\text{QMC}}$. In the CT-HYB calculations, we used the density-density component of the general Coulomb interaction matrix [see Eq.~(\ref{eq:U_Slater})].\label{fig:ct_spectra}}
\end{figure}

\begin{figure}[t]
\centering
\includegraphics[width=\linewidth]{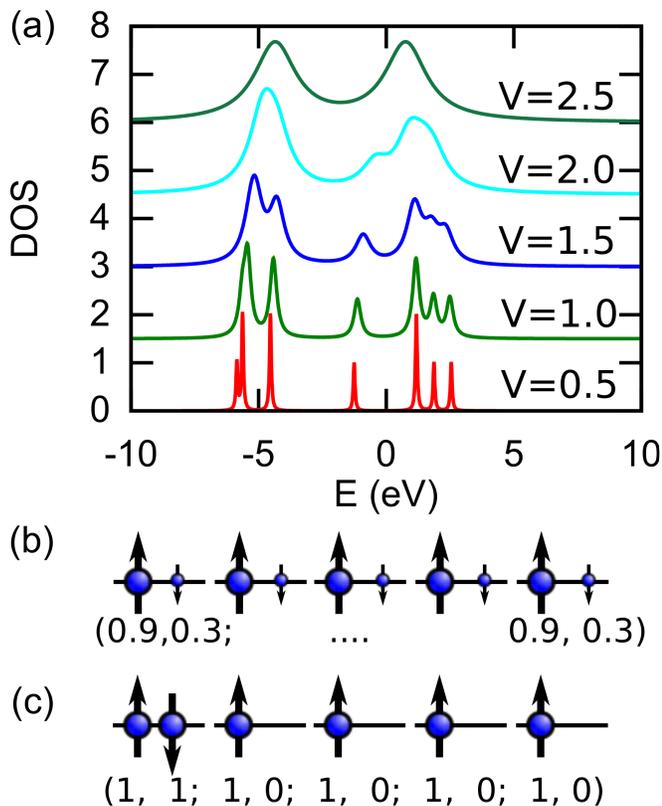}
\caption{(Color online) (a) Spectral functions at different hybridization strengths $V$ obtained in the HF approximation. The results are for a density-density type Coulomb interaction matrix. (b,c) Illustrations of the HF ground state impurity occupations at $V=2.5$ and in the atomic limit $V=0.0$.\label{fig:hf_spectra}}
\end{figure}

\begin{figure}[t]
\centering
\includegraphics[width=\linewidth]{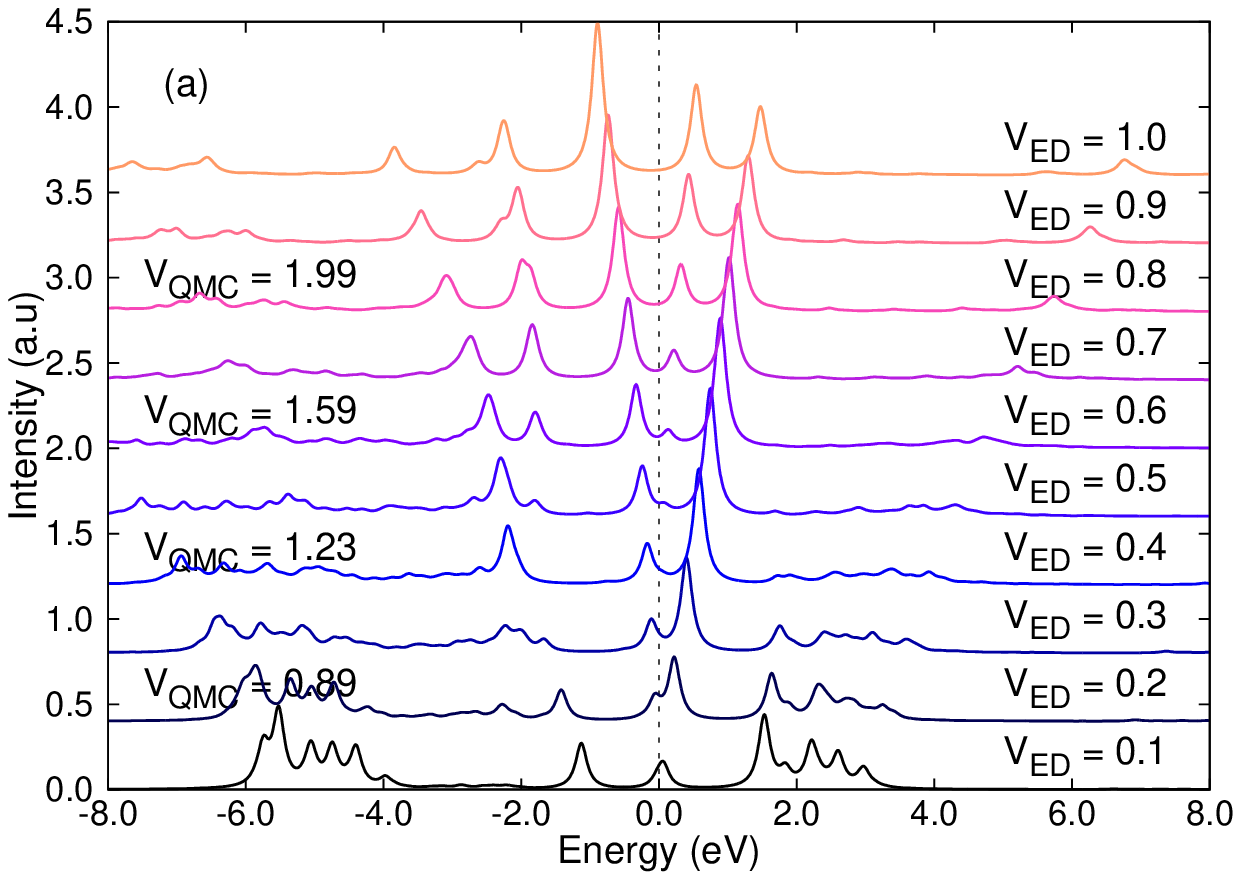}
\includegraphics[width=\linewidth]{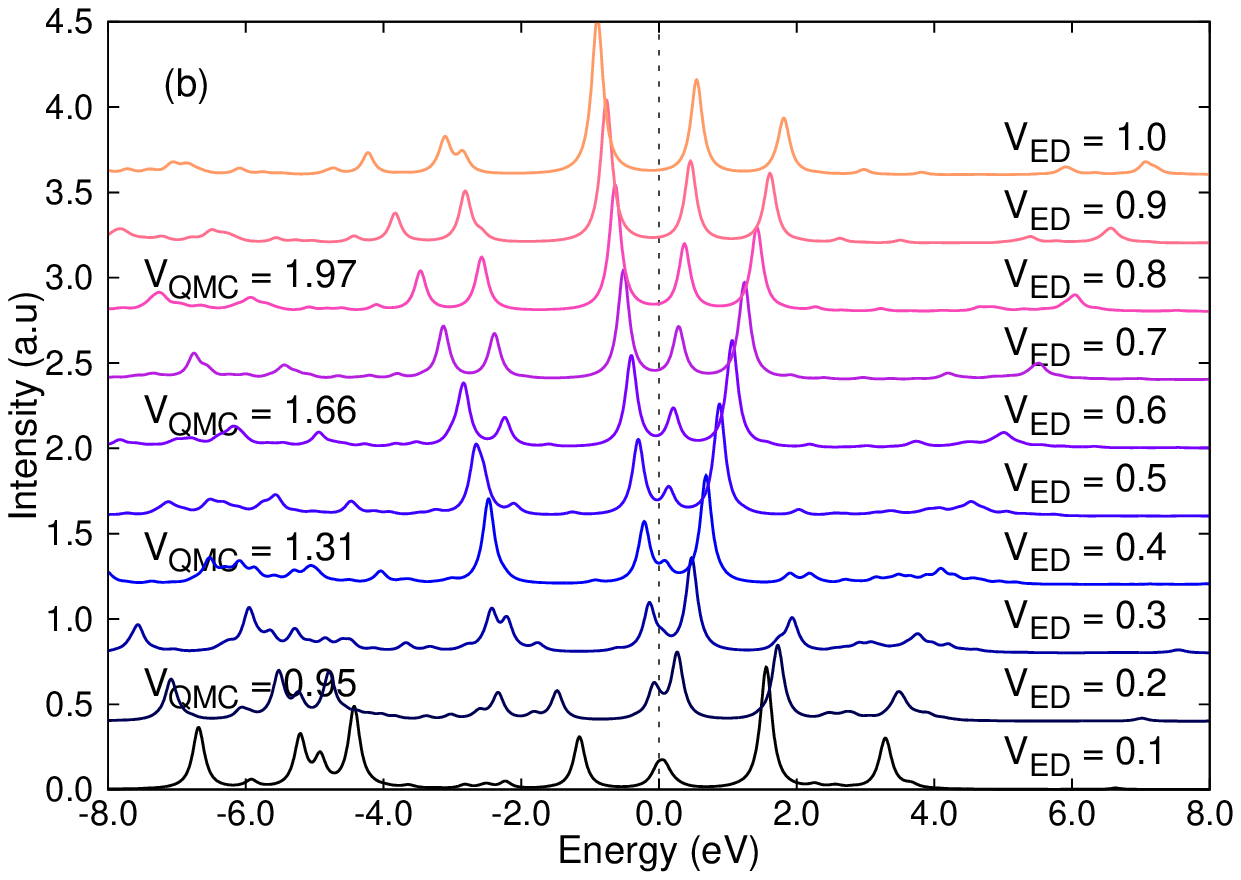}
\caption{(Color online) Calculated spectral functions at different hybridization strengths $V_{\text{ED}}$. For some selected curves, the corresponding $V_{\text{QMC}}$ is given. (a) Results for the density-density part of the rotationally invariant interaction matrix. (b) Fully rotationally invariant interaction matrix. \label{fig:ed_spectra}}
\end{figure}

\begin{figure*}[t]
\centering
\includegraphics[width=\linewidth]{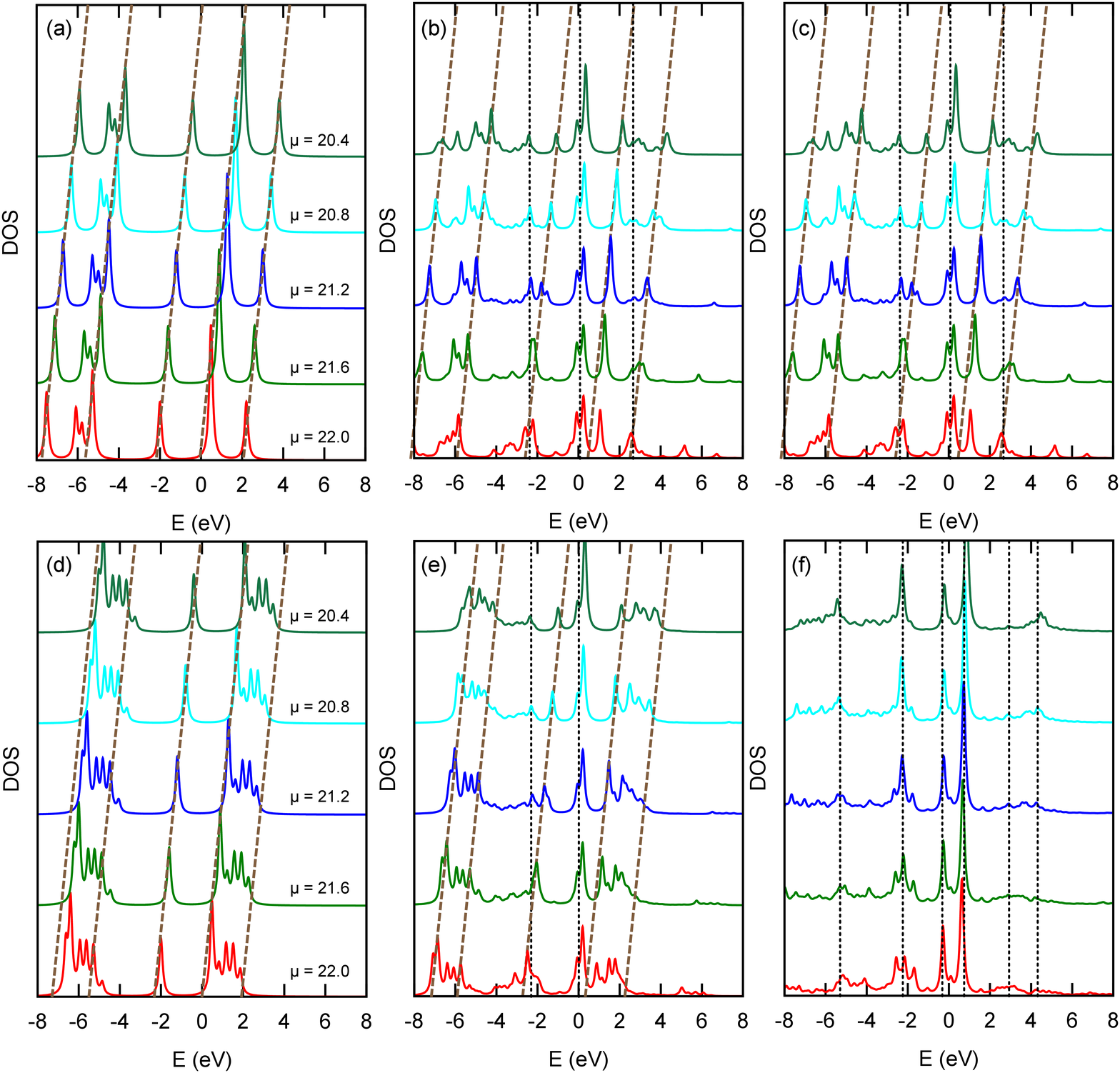}
\caption{(Color online) Dependence of the ED spectral functions on the chemical potential $\mu$. Top panels: Full rotationally invariant Coulomb matrix. Bottom panels: Density-density interaction. Spectral functions for the $V_{\text{ED}}=0.0$ atomic limit (a,d), $V_{\text{ED}}$ = 0.2 (b,e), and $V_{\text{ED}}$ = 0.5 (c,f) are shown. Peaks shifting with the chemical potential according to ${\partial \omega}/{\partial \mu}\sim 1$ (brown dashed lines) can be distinguished from the peaks displaying a small $\mu$-dependence $|{\partial \omega}/{\partial \mu}|\ll 1$ (black dotted lines). \label{fig:multi}}
\end{figure*}

The central goal of this section is to study the competition of atomic charging and multiplet features with hybridization effects. Both manifest themselves in the one-particle spectral function which can be measured by photoemission \cite{PhysRevLett.110.186404} or scanning tunneling spectroscopy.\cite{Wenderoth.Fe.Kondo.2011,PhysRevLett.106.037205} In particular we study the emergence of quasiparticle peaks and the robustness of atomic multiplets with hybridization.
 
We first discuss the evolution of the CT-HYB spectral function upon increasing the hybridization of the impurity with the bath, as shown in Fig.~\ref{fig:ct_spectra}. For sufficiently small hybridizations, $V_{\text{QMC}}\sim 0.1$, the CT-HYB spectra reveal the multiplet structure of isolated atoms and solely consist of upper and lower ``Hubbard bands", as expected. Starting from $V_{\text{QMC}}\approx 0.5$, however, a quasiparticle peak close to the Fermi level emerges, which gains spectral weight with increasing hybridization. At $V_{\text{QMC}} \sim 2.0$ the quasiparticle peak is so broad that it can not be distinguished from the lowest multiplet of the upper Hubbard band or the highest multiplet of the lower Hubbard band any more. The multiplet feature between $-5$ and $-6$ eV slightly shifts and broadens but otherwise survives the increasing hybridization. In addition, a spectral satellite develops, around $-2.4$~eV upon increasing the hybridization.

To gain some insights into the physical origin of this evolution we compare to HF (see Fig.~\ref{fig:hf_spectra}) and ED calculations (see Fig.~\ref{fig:ed_spectra}). At small hybridization, the HF spectra show atomic ionization peaks and qualitatively agree with the ED and CT-HYB spectra. In this case, we find a symmetry broken HF ground state, which corresponds to an $N=6$, $L=2$, $S=2$ atomic configuration. It is spin- and orbitally polarized, as illustrated in Fig.~\ref{fig:hf_spectra}(c). Upon increasing the hybridization, the orbital polarization decreases and vanishes between $V=2.0$ and $2.5$ in the HF model, while the spin polarization persists, c.f. Fig.~\ref{fig:hf_spectra}(b). Correspondingly, the HF spectral function evolves from several multiplet peaks at low hybridization to a two peaks structure at $V>2.0$. The spectral peak around $-5.0$~eV is associated with the emission of a majority spin electron from the impurity, while the peak around the Fermi level stems from the minority electrons. The robust spectral feature in the CT-HYB calculations between $-5.0$ and $-6.0$~eV can thus be traced back to spin-exchange splitting and the emission of (instantaneous) majority spin electrons from the impurity. After this emission, the impurity is in a low-spin state. Also the minority electron spectral features from the HF approximation find their counterparts in the CT-HYB calculations both in the case of small hybridization $V\lesssim 0.5$ as well as for the stronger hybridizations under investigation ($V \sim 2.0$). In the case of strong hybridization, the spectral features around $-6.0$ and $0.0$ eV can be explained in terms of a HF picture.

There are, however, several features in the CT-HYB spectra, which do not have a counterpart in the HF spectra. First, there is no quasiparticle peak at the Fermi level in the HF calculations at intermediate coupling $0.5<V<1.5$. Second, in contrast to HF, the CT-HYB calculations reveal the formation of a satellite peak at $-3.0$~eV~$<\omega<-2.0$~eV for hybridizations exceeding $V \gtrsim 1.0$. Both features are therefore likely due to dynamic correlation effects and should be related to multi-determinant final or initial states. Here, ED calculations can provide useful insights since they can capture multi-determinant effects and yield spectral functions without the need of analytical continuation.

Figure~\ref{fig:ed_spectra} shows the ED spectra for bath couplings increasing from $V_{\text{ED}}=0.1$ to $V_{\text{ED}}=1.0$. Obviously, the discrete bath in ED leads to differences between the ED and the CT-HYB spectra. However, both the occurrence of a quasiparticle feature pinned to the Fermi level at intermediate couping as well as the satellite peak between $-2.0$ and $-3.0$~eV are also observable in the ED spectra. Clearly, neither the quasiparticle feature nor the satellite peak between $-2.0$ and $-3.0$~eV have a counterpart in the electronic spectra for the $N=6$ atomic limit [see Fig.~\ref{fig:ct_v01_ed_v00}(a)]. Of course, all peaks in the atomic spectra correspond to ionization processes where the electron number at the atomic site changes by $\Delta N=\pm 1$. As soon as there is hybridization, the electron number at the atomic site fluctuates and spectral peaks can result from processes, where the impurity occupation in the initial and final states remains essentially constant, i.e., $|\Delta N| \ll 1$.

In fact, the $\Delta N$ associated with a given spectral feature can be easily determined by analyzing the variation of the peak energy with the impurity on-site energies $\epsilon_\alpha$ or equivalently the chemical potential $\mu$. For the initial state $\ket{i}$ (e.g. the ground state of the system) the energy eigenvalue $E_i$ changes with $\mu$ according to
\begin{align}
-\frac{\partial E_i}{\partial \mu}&=-\bra{i}\partial H_{\text{loc}}/\partial \mu\ket{i}=\sum_{\alpha}\bra{i} d^{\dagger}_{\alpha}d_{\alpha}\ket{i}=N_i,
\label{eq:eps_a_variation}
\end{align}
where $N_i$ is the average number of electrons at the impurity site in the initial state. Similarly, for every final state the change of its energy eigenvalue $E_f$ upon variation of the chemical potential $-{\partial E_f}/{\partial \mu}=N_f$ is given by the average impurity occupancy $N_f$ in that particular final state. Thus the energy $\omega=\pm|E_i-E_f|$ of each peak in the spectral function will shift according to ${\partial \omega}/{\partial \mu}=\pm|N_i-N_f|=\pm |\Delta N|$.

Fig.~\ref{fig:multi} shows the variation of the ED spectra for different hybridization strengths with the chemical potential $\mu$. In the atomic limit, all spectral peaks shift according to $|{\partial \omega}/{\partial \mu}|=1$, as it must be. As soon as there is hopping $V$ to the bath orbitals, the additional spectral features at the Fermi level and around $-2.5$~eV appear. These shift much less upon variation of the chemical potential, $|{\partial \omega}/{\partial \mu}|\ll 1$, which suggests that the impurity occupancy difference between initial and final states is nearly zero, here: $\Delta N\approx 0$. The final states associated with the quasiparticle peak have a large contribution from the $N=6$ impurity ground state multiplet (i.e. spin and orbital angular momenta $S=2, L=2$). For the free atom, the lowest multiplet excitations within the $N=6$ subspace lead to spin $S=1$ states with different total orbital angular momenta $L=5,1,3,4$ (for the rotationally invariant interaction) and excitation energies between $2.2$ and $2.7$~eV. The final states of the satellite features in the energy range between $\omega=-2.0$~eV and $-3.0$~eV are largely derived from these $N=6$, $S=1$ multiplets (with a hole in the bath orbitals). We note that the satellite features are observable in our CT-HYB and ED calculations up to the highest hybridization strengths under investigation. They persist even strong charge fluctuations $\langle \Delta N^2\rangle \sim 1$.

\subsection{Spin freezing and non-Fermi-liquid behavior}
\begin{figure*}[t]
\includegraphics[width=\columnwidth]{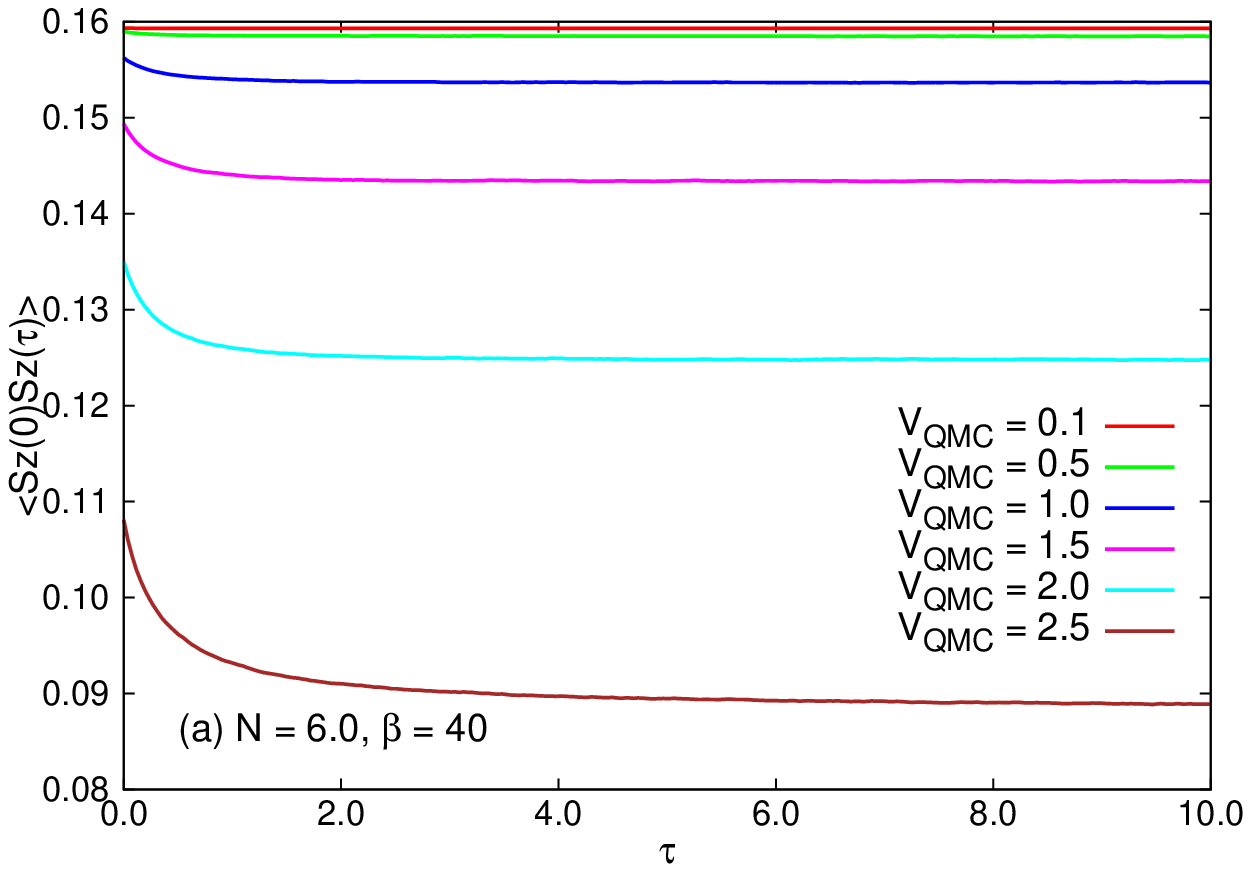}
\includegraphics[width=\columnwidth]{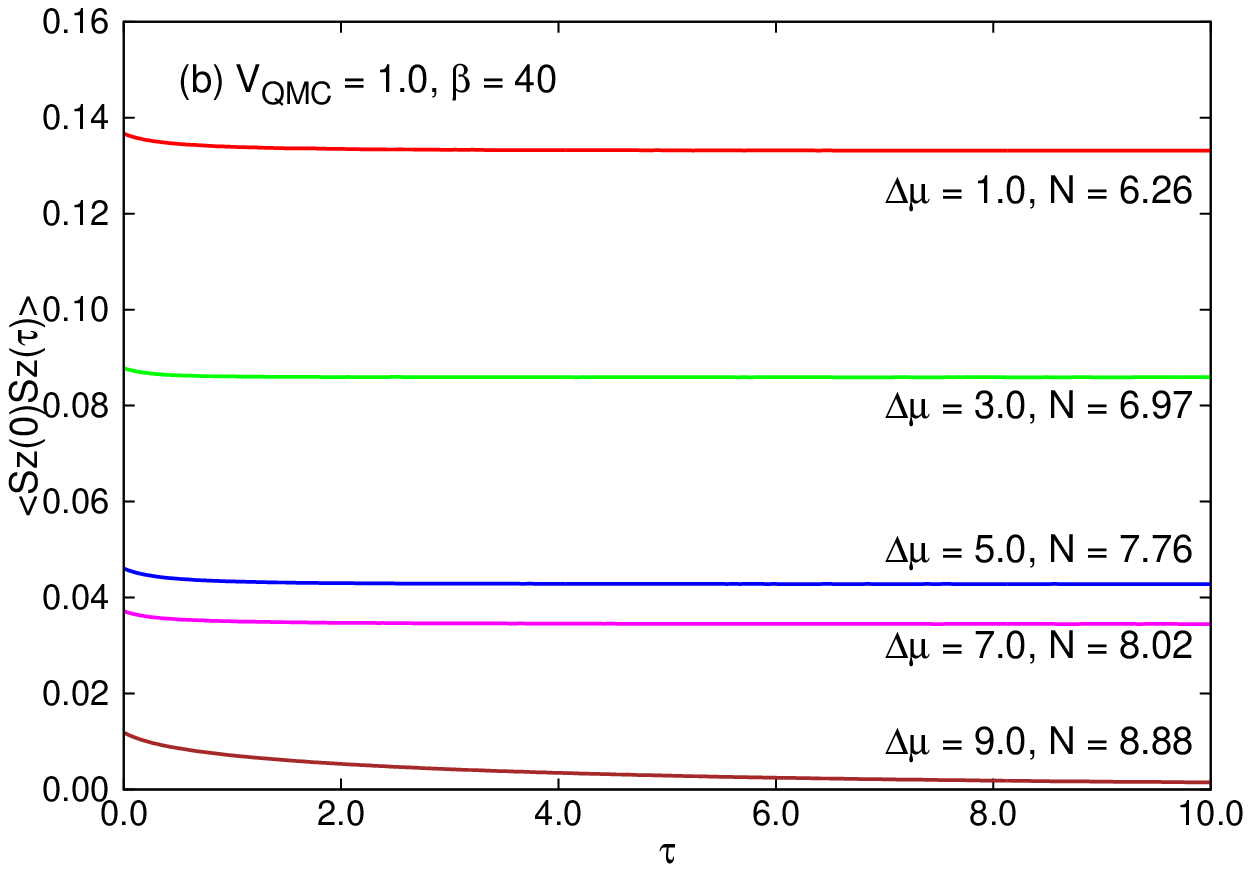}
\includegraphics[width=\columnwidth]{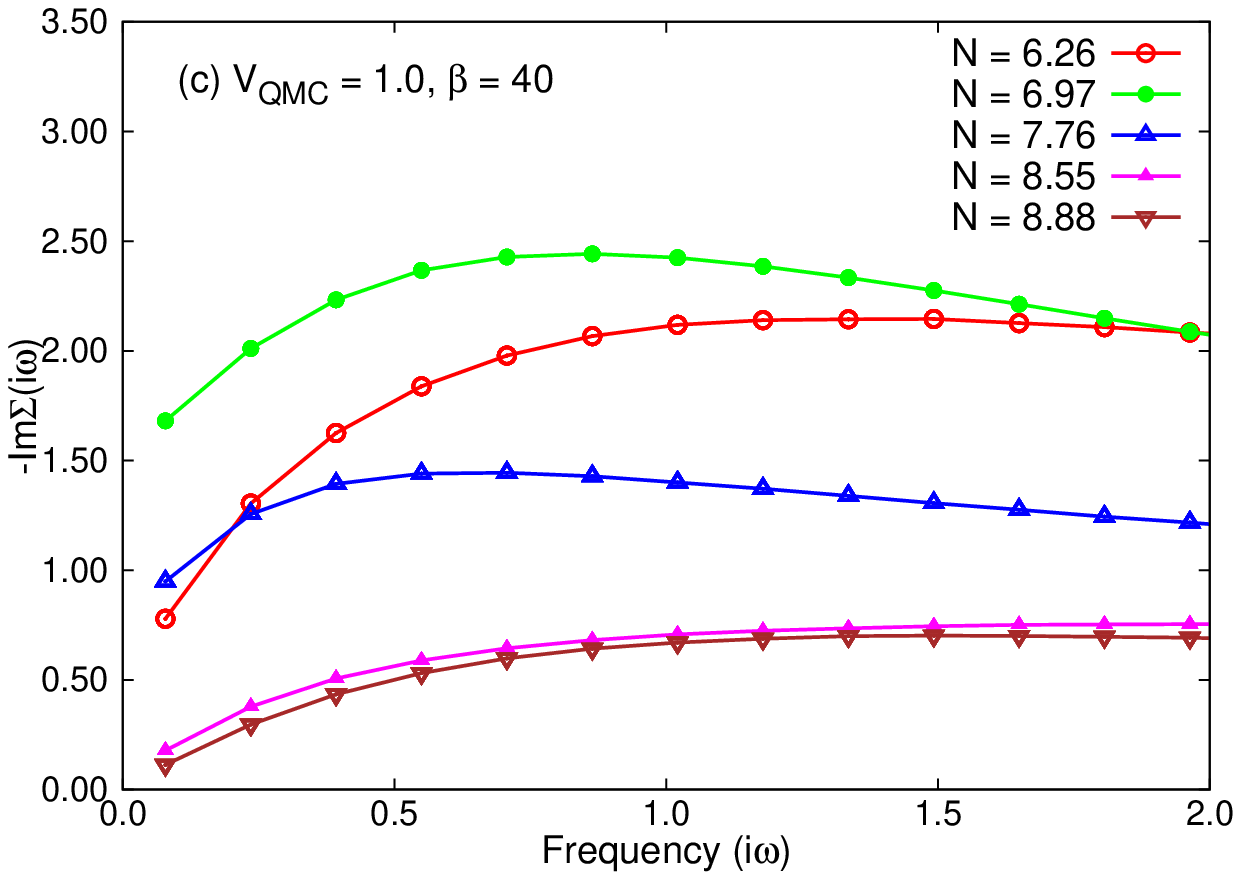}
\includegraphics[width=\columnwidth]{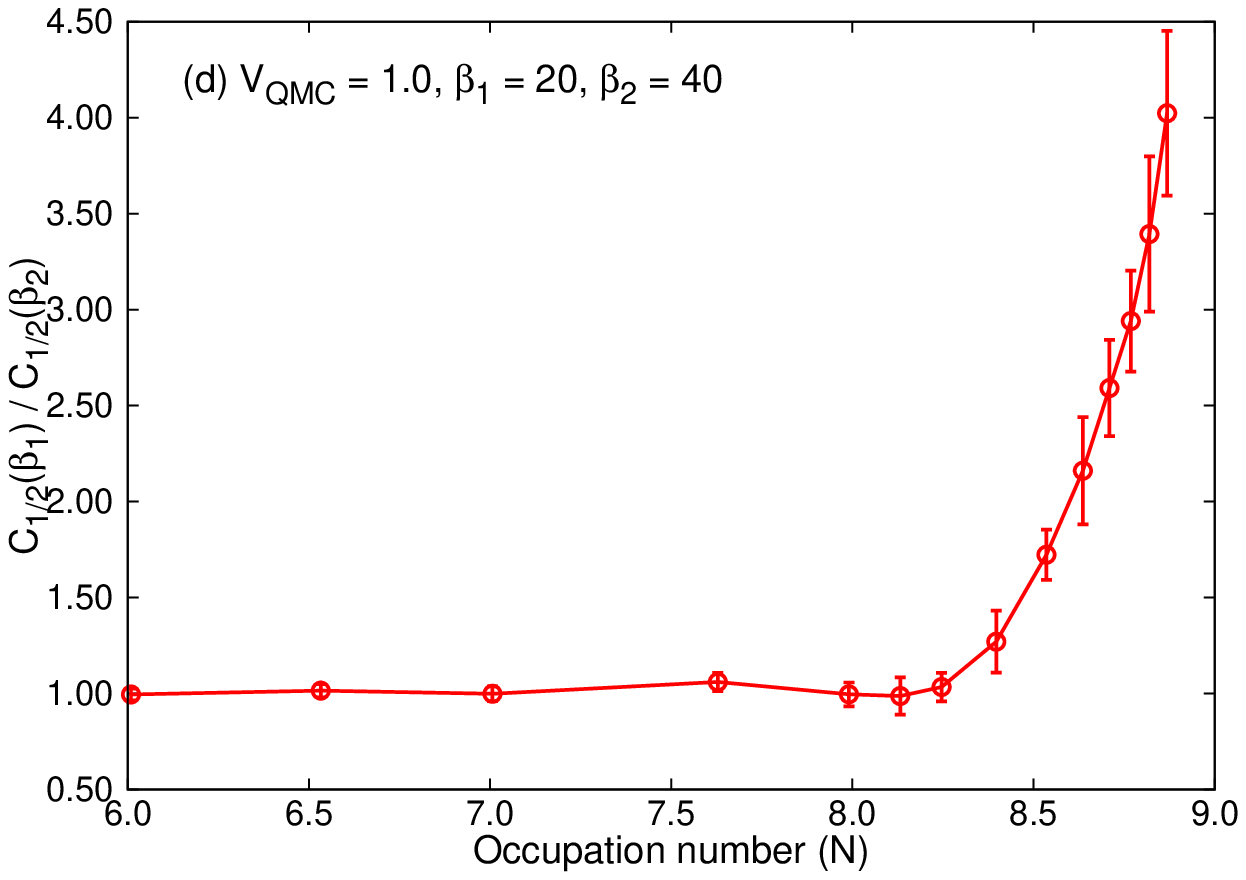}
\caption{(Color online) Dynamic spin-spin correlation function $\langle S_z(\tau)S_z(0)\rangle$ and self-energy function of the five-orbital Anderson impurity model obtained in the CT-HYB calculations. Here, $S_z=\frac{1}{5}\sum_\alpha \frac{1}{2}(n_{\alpha\uparrow}-n_{\alpha\downarrow})$ is the electron spin-density. (a) Spin-spin correlation functions for different hybridization strengths $V_{\text{QMC}}$ ($N = 6$, $\beta = 40$). (b) Spin-spin correlation functions for different electron occupations ($V_{\text{QMC}} = 1.0$, $\beta = 40$). (c) Imaginary part of low-frequency self-energy function -Im$\Sigma(i\omega_n)$ for different electron occupations ($V_{\text{QMC}} = 1.0$, $\beta = 40$). (d) $C_{1/2}(\beta_1)/C_{1/2}(\beta_2)$. Here $C_{1/2}(\beta)$ denotes the $\tau = \beta/2$ value of the spin-spin correlation function: $C_{1/2}(\beta) = \langle S_z(\beta/2)S_z(0)\rangle$ ($V_{\text{QMC}} = 1.0$, $\beta_1 = 20$, and $\beta_2 = 40$). \label{fig:SzSz}}
\end{figure*}

In a broader context it is interesting to note that characteristic correlation effects arising from Hund's coupling have been identified in multiorbital lattice models.\cite{Werner2008,deMedici2011,Haule2009,Liebsch2010} Independent of the details of the interaction matrix, in a certain doping range away from half-filling, disordered local moments appear in the metallic phase, and the Fermi-liquid coherence temperature becomes very low. This phenomenon, which has been dubbed ``spin freezing"\cite{Werner2008} or ``Hund's metal"\cite{deMedici2011} is believed to explain the unusual properties of important classes of correlated materials, including ruthenates\cite{Werner2008, Georges2013} and iron pnictides.\cite{Haule2009, Liebsch2010,Werner2012} 

The itinerant atomic magnetism observed in our study of the five-orbital Anderson impurity model corresponds to a single atom realization of this spin frozen metallic state. This can be most directly seen from the imaginary time dynamic spin-spin correlation function $\langle S_z(\tau)S_z(0)\rangle$ of the five-orbital impurity. In Fig.~\ref{fig:SzSz}(a) we plot the spin-spin correlation functions for different hybridization strengths $V_{\text{QMC}}$ and electron occupation $N = 6$. We find that $\langle S_z(\tau)S_z(0)\rangle$ does not decay to zero at large imaginary times $\tau$, which indicates ``spin freezing".\cite{Werner2008} We also considered the filling dependence of the spin-spin correlation function [Fig.~\ref{fig:SzSz}(b)]. As the electron occupation increases, the local magnetic moment decreases, but the spin-freezing phenomenon persists up to a filling of about $N=8$. When $N$ increased to $9.0$, the value of $\langle S_z(\tau)S_z(0)\rangle$ at large $\tau$ approaches zero, which indicates a crossover to a Fermi-liquid metal state.

To study the crossover from the non-Fermi-liquid metallic state with frozen local moments to a Fermi-liquid metal we follow the procedure outlined in Ref.~\onlinecite{Werner2008}. Let us define $C_{1/2}(\beta)$ as the value of the spin-spin correlation function at the mid-point of the imaginary-time interval: $C_{1/2}(\beta) = \langle S_z(\beta/2)S_z(0)\rangle$. The Fermi-liquid or non-Fermi-liquid behavior can be seen in the temperature dependence of this quantity, i.e. by plotting the ratio $C_{1/2}(\beta_1)/C_{1/2}(\beta_2)$ as a function of filling (here we choose $\beta_1 = 20$ and $\beta_2 = 40.0$). The results are shown in Fig.~\ref{fig:SzSz}(d). In a Fermi-liquid state, $C_{1/2}(\beta) \propto 1/\beta^2$, while in the frozen moment phase $C_{1/2}(\beta)$ becomes temperature independent at sufficiently low $T$. Fig.~\ref{fig:SzSz}(d) clearly shows the crossover from the value $1$ expected in the spin frozen phase ($N \lesssim 8.0$) to the value 4 expected in the Fermi-liquid metallic phase ($N \gtrsim 9.0$). The large error bars at large electron occupation are mainly caused by the tiny values of the spin-spin correlator at $\tau=\beta/2$. 

While the existence of local moments in the spin frozen state can be well understood in a static mean-field picture (see Fig.~\ref{fig:hf_spectra}), the low energy excitations differ from standard Fermi-liquid quasiparticles. The latter exhibit vanishing decay rates at low energies and low temperatures, which implies a vanishing imaginary part of the Matsubara self-energy, $\text{Im}\Sigma(i\omega_n\to 0)\to 0$. In our CT-HYB calculations we find, however, finite scattering rates $\text{Im}\Sigma(i\omega_n\to 0)\to \Gamma$ due to the local impurity moment. Only at very low temperatures a Fermi-liquid metal with small quasiparticle weight might emerge. In Fig.~\ref{fig:SzSz}(c) we plot the Matsubara self-energy for different fillings. For $N < 8$, Im$\Sigma(i\omega_n)$ at low-energies shows an obvious non-Fermi-liquid character. However, when $N$ increases to 9, the scattering rate goes to zero, and the low-frequency Im$\Sigma(i\omega_n)$ exhibits roughly a linear behaviour with frequency, which means that the model is in the vicinity of a Fermi-liquid metallic state. 

\section{conclusion\label{sec:conclusion}}
In this work, we analyzed the evolution of (thermodynamic) ground state and excited states properties of the five-orbital Anderson impurity model from the atomic limit to hybridization strengths which correspond to typical metallic environments. A numerically exact CT-HYB approach was combined with HF and ED calculations to pinpoint the physical mechanisms affecting the impurity spectral function, which can be measured by single particle spectroscopies like photoemission or scanning tunneling spectroscopy.

Since ED works with a discretized bath (in our case one bath orbital per impurity orbital directly at the Fermi level), it is \emph{a priori} unclear how the original hybridization strength $V_{\text{QMC}}$ and the bath coupling in ED, $V_{\text{ED}}$, should be matched. We showed that choosing $V_{\text{ED}}$ such that it yields the same average charge fluctuations as the full model is not only very natural, but also leads to a good agreement between the ED and the CT-HYB spectral functions (both for multiplet features and many body satellite peaks) over an energy range of several eV.

We found that multiplet features are observable in the entire range of hybridization strengths under investigation ($0.0 \leq V_{\text{QMC}} \leq 2.0$), even if the quasiparticle peak is so wide at $V_{\text{QMC}}=2.0$ that it cannot be distinguished from all Hubbard band features. The impurity magnetic moment is stabilized by the Hund's coupling $J$ and can persist under strong charge fluctuations. A single five-orbital impurity can thus realize a situation which is very similar to itinerant bulk magnets or bulk Hund's metals.\cite{Werner2008,deMedici2011,Haule2009,Liebsch2010,Georges2013} We note that this single impurity ``itinerant" magnetism is a genuine multiorbital effect, which cannot be realized in the single orbital Anderson model. In the latter model, solely the Hubbard $U$ is responsible for the formation of magnetic moments and strong charge fluctuations exclude sizable magnetic moments. Consequently, itinerant behavior and Kondo physics mutually exclude each other in the single orbital model but not necessarily in more realistic five-orbital models of $d$-electron systems. On the contrary, genuine many electron features like the multiplet excitation satellites between $-2.0$ and $-3.0$~eV are largely enhanced in the case of considerable charge fluctuations in the multiorbital model. 

Fe impurities in noble metal environments such as bulk Au or Ag surfaces are classical examples of multi-orbital quantum impurity systems\cite{deHaas19341115} and are still widely studied.\cite{PhysRevLett.97.226803,PhysRevLett.97.226804,PhysRevLett.102.056802,PhysRevLett.110.186404,Wenderoth.Fe.Kondo.2011,PhysRevLett.106.037205,PhysRevB.84.113112,PhysRevB.88.075146} Fe in Au and Ag displays a low temperature resistance minimum, which has been interpreted in terms of Kondo physics:\cite{HewsonBook} an impurity \textit{spin} couples antiferromagnetically to the conduction electrons and is screened around the Kondo temperature, which is here on the order of 5\ K - 40\ K.\cite{PhysRevLett.97.226803,PhysRevLett.97.226804,Wenderoth.Fe.Kondo.2011} Scaling analyses of weak localization quantum transport experiments \cite{PhysRevLett.97.226803,PhysRevLett.97.226804} in combination with NRG calculations \cite{PhysRevLett.102.056802,PhysRevB.88.075146} showed that Fe in bulk noble metals appears to realize an effective spin of $S\geq 3/2$. The microscopic nature of this spin has remained however unclear. Photoemission spectroscopy (PES) probes energy scales from several 10~meV to a few eV and has revealed a several 100~meV broad quasiparticle peak near the Fermi level, as well as sizable charge fluctuations still coexisting with exchange split multiplet features for Fe impurities on a Ag surface.\cite{PhysRevLett.110.186404} These PES experiments and also inelastic scanning tunneling spectroscopy\cite{PhysRevLett.106.037205} (STS) experiments suggest a rather itinerant behavior of Fe impurities in transition metal hosts. 

In agreement with PES of Fe on a Ag surface,\cite{PhysRevLett.110.186404} our calculated spectra show multiplet features persisting under sizable charge fluctuations. At the same time we find a ``spin-freezing" behavior in the spin-spin correlation functions. Since the spin-frozen state crosses over into a Fermi-liquid state at very low temperatures, the concept of itinerant single atom magnets and spin-freezing put forward here may reconcile seemingly contradictory quantum transport, PES and STS experiments. Future investigations involving more realistic hybridization functions, crystal fields and spin-orbit coupling will be useful to clarify this hypothesis further.

The electronic structure of multiorbital impurities is controlled by a complex interplay of charge, spin and orbital fluctuations, which can be very sensitive to the particular form of the Coulomb interaction matrix assumed in the model.\cite{PhysRevB.80.155132} It is, however, not \emph{a priori} clear how particular ground state properties or spectral features are affected by a certain approximation of the Coulomb interaction matrix. Taken together, our ED and CT-HYB calculations show that any approximation to the Coulomb vertex should be constructed such that it leaves the amount of charge fluctuations in the ground state unaffected. For instance, the fully rotationally invariant Coulomb vertex, Eq.~(\ref{eq:U_Slater}), and its density-density part lead to very similar charge fluctuations and indeed very similar spectra (up to some degeneracy lifting) over a range of several eV. However, the spectra and the amount of charge fluctuations derived from impurities with the Slater-Kanamori ``$U-3J$" interaction, Eq.~(\ref{eq:hint_sk}), are qualitatively different. The ``$U-3J$" interaction does not reproduce the Mott gap (effective charging energies) of the fully rotationally invariant interaction, while the density-density part of Eq.~(\ref{eq:U_Slater}) does.

The neglect of non-density-density terms becomes problematic wherever a precise description of local degeneracies is crucial. This is not so much the case for higher energy spectral features but clearly for low temperature or low energy features. We anticipate that quantities like Kondo temperatures, the shapes of Kondo resonances or also magnetic anisotropies in rotation symmetry broken structures can be very sensitive to the non-density-density terms in the Coulomb vertex.

\begin{acknowledgments}
We acknowledge financial support from SNF Grant No.~200021\_140648 and from the DFG via FOR 1346. TOW thanks J. Kolorenc (FZU, Prague) for providing his ED code as well as A. Lichtenstein and G. Czycholl for useful discussions.
\end{acknowledgments}

\bibliography{test}
\end{document}